\documentclass[onecolumn,prb]{revtex4}
\usepackage{graphicx,psfrag,color}
\usepackage{slashed,amssymb,amsmath,dsfont}
\newcommand{\nn}{\nonumber}

\begin{document}
 
\title{
Diffusive transport in the lowest Landau level of disordered 2d semimetals:
the mean-square-displacement approach
}

\author{Andreas Sinner$^{1,2}$, and Gregor Tkachov$^{1}$,}
\affiliation{
\mbox{$^{1}$ Institut f\"ur Physik, Universit\"at Augsburg, D-86135 Augsburg, Germany}\\
\mbox{$^{2}$ Instytut Fizyki, Uniwersytet Opolski, PL-45-052 Opole, Poland}
}

\date{\today}

\begin{abstract}
We study the electronic transport in the lowest Landau level of disordered two-dimensional semimetals placed in a homogeneous perpendicular magnetic field.
The material system is modeled by the Bernevig-Hughes-Zhang Hamiltonian, which has zero energy Landau modes due to the material's intrinsic Berry curvature. 
These turn out to be crucially important for the density of states and the static conductivity of the disordered system. 
We develop an analytical approach to the diffusion and conductivity based on a self-consistent equation of motion for the mean squared displacement. 
The obtained value of the zero mode conductivity is close to the conductivity of disordered Dirac electrons without magnetic fields, 
which have zero energy points in the spectrum as well.
Our analysis is applicable in a broader context of disordered two-dimensional electron gases in strong magnetic fields.
\end{abstract}

\maketitle

\section{Introduction}
\label{Sec:Intro}

The subjects of low dimensional semimetals and related topics of topological states of matter and spintronics have witnessed an enormous boom during the last decades~[\onlinecite{{Guinea2009,Zhang2008,Hasan2010,Kotov2012,Qi2011,Tkachov2020,Avsar2020}}]. 
In magnetic fields, the quantum mechanics of charge carriers in these materials is governed by a spectacular interplay of the intrinsic and magnetic-field-induced Berry curvatures. 
Several aspects of this fascinating physics remain widely untouched, though. For instance, relative little is known about the role of disorder and its interplay with the magnetic field. 
The overall progress in this area has been slow, not least because of the technical challenges, considerable even by the standards of the community~[\onlinecite{Ando1974,Wegner1983,Brezin1984,Hikami1984a,Hikami1984b}].
There is a number of issues which make the disordered electrons in the homogeneous perpendicular magnetic field look differently as compared to the situation without a magnetic field. 
Due to the freedom of the gauge choice, the problem can be approached in a number of ways, which differ very much in details. For instance, 
the choice of the central gauge has the advantage that the solutions of the Schr\"odinger equation are states localized in the space. 
Therefore one can do computations in the position space in an exact manner.
The envisaged problem is notoriously difficult because the model lacks a small expansion parameter~[\onlinecite{Aoki1987}].
This inevitably leads to divergent expansion series. A powerful method to keep such divergences under control is the renormalization group. 
In the past, our understanding of the physics of disordered metals and semiconductors profited vastly from the various combinations of variational 
and perturbative techniques with the renormalization group, c.f. Refs.~[\onlinecite{Abrahams1979,Gorkov1979,Hikami1980,Wegner1979a,Wegner1980,Hikami1981}] 
and Refs.~[\onlinecite{Tkachov2015,Sinner2014,Sinner2016}]. 

However, in the central gauge picture, there is literary no continuous variable to be sliced off by iterations in order to obtain the renormalization group equations.
Of course one can use a different gauge, which allows for description in terms of states localized in one direction and propagating in the other. 
The price to pay is the loss of exactness, which is too costly to give up. 
In this paper we develop a diagrammatic approach to the conductivity of the two-dimensional disordered electron gas in strong magnetic field in central gauge picture. 
While these series can still be wrapped up exactly for the single-particle propagators, 
as it was impressively demonstrated by F. Wegner in Ref.~[\onlinecite{Wegner1983}], additional technical issues make elusive every attempt of applying these techniques to the 
two-particles propagators. The available divergent series cannot be directly plugged into the Kubo formula without some not a-priory obvious sort of regularization or resummation.
Hence, the usual way to approach the conductivity is via the Einstein relation and correspondingly via the notion of diffusion~[\onlinecite{Goldenfeld,Huang,Chaikin}].
Because the corresponding statistical averages require normalization with respect to the vacuum fluctuations~[\onlinecite{Ziegler2012}], 
this provides a tool of estimating the measurable quantities by means of some kind of analytical continuation~[\onlinecite{Hikami1984a,Hikami1984b,Chakravarty1986,Hikami1993}].

Our approach may not differ much from the others in relaying on perturbative expansions for the two-particles propagator. 
It too requires as much information from the perturbative expansion as possible and so, we perform the exact computations of perturbative series to the practical limits of doable. 
Moreover, we even go beyond that. We classify diagrammatic channels according to their behavior on large scales in the position space and identify the dominant ones.
In principle this gives a hint towards the exact asymptotics of the correlation functions. 
What is different in our work is the way we approach the mean squared displacement and from this the diffusion coefficient. 
Instead of resumming the perturbative series, our departing point is the individual behavior in time of each element of the perturbative series. 
It turns out that the behavior at larger time is dominated by the higher order elements and tends towards a stationary state. 
On the sublaying time scales though, there is a large region with linear time dependence, characteristic of the diffusion.
To approach this regime we propose a self-consistent equation of motion for the mean squared displacement and extract the diffusion coefficient from there.
With the obtained diffusion coefficient and density of states we find via the Einstein relation a universal expression for the static conductivity in the lowest Landau level.

For concreteness we chose the Hamiltonian proposed by Bernevig, Hughes and Zhang 
in the context of the quantum spin Hall effect~[\onlinecite{Zhang2006,Zhang2008}]. Its spectral properties are well known and thoroughly studied in the cited literature. 
The distinctive feature of the BHZ-Hamiltonian in the lowest Landau level is the appearance of states with zero energy at a certain strength of the magnetic field.
These Landau zero modes prove crucial to the transport in presence of disorder. Both the density of states and the conductivity obtained from the Einstein relation
are peaked around these modes. In all the system becomes metallic within a parameter window around them, which becomes broader but also less expressed with increasing disorder. 
The numerical value exactly at a zero point is close to that of Dirac electrons in random potentials.

The paper is organized as follows: In Section~\ref{Sec:Ham} we specify the microscopic BZH-Hamiltonian which we use for concrete discussions. 
We discuss the specifics of its spectrum and eigenstates, and introduce the corresponding Green's function in Section~\ref{Sec:LLLGF}. 
In Section~\ref{Sec:Dress} we discuss the effect of the different types of disorder and the density of states. 
In Section~\ref{Sec:OptCon} we discuss the mean-squared displacement and its relation to the diffusion. 
We present an explicit evaluation of the mean-squared displacement to any finite order in perturbative expansion. 
For this we determine an exact asymptotic expression for the two-particles propagators.
In Section~\ref{Sec:EMMSD} we propose a self-consistent equation of motion for the mean-squared displacement and 
extract the diffusion coefficient from that. This gives us an access to the static conductivity. 
Lengthy auxiliary calculation pieces are moved into the Appendix.

\section{The model Hamiltonian}
\label{Sec:Ham}

In the absence of magnetic field the BHZ-Hamiltonian reads 
\begin{equation}
\label{eq:BHZH}
H = \Delta^{}_0\Sigma^{}_{03} - \epsilon^{}_0\Sigma^{}_{00} - \left({\cal C}^{}_+ T^{}_- + {\cal C}^{}_- T^{}_+\right)\nabla^{}_+\nabla^{}_- - iv\left({\cal D}^{}_+\nabla^{}_- + {\cal D}^{}_-\nabla^{}_+\right).
\end{equation}
the Hamiltonian is shown in its explicit form  in Appendix~\ref{app:Ham}. The shorthands used here are  
$\displaystyle \nabla^{}_{\pm}  = \partial^{}_x \pm i\partial^{}_y, \;\; \nabla^{}_{+}\nabla^{}_{-} = \partial^2_x + \partial^2_y, \;\; {T^{}_\pm} = \frac{1}{2m}\pm B^{}_0$. 
The $4\times4$ matrix body of the Hamiltonian is spanned by some of the 16 Dirac matrices $\displaystyle \Sigma^{}_{ab} = \sigma^{}_a\otimes\sigma^{}_b$, first index referring to the spin space,
$a,b = 0,1,2,3$, with $\sigma^{}_{a=1,2,3}$ denoting the Pauli matrices in their usual representation and $\sigma^{}_{a=0}$ being the two-dimensional unity matrix. 
With this the matrices ${\cal C}^{}_\pm$ and ${\cal D}^{}_\pm$ read
\begin{equation}
{\cal C}^{}_\pm = \frac{1}{2}\left[\Sigma^{}_{00}\pm\Sigma^{}_{03} \right], \;\; {\rm and} \;\;\, {\cal D}^{}_\pm = \frac{1}{2}\left[\Sigma^{}_{01}\pm i\Sigma^{}_{02}\right].
\end{equation}
The band gap $\Delta^{}_0$ is supposed to be much smaller than the typical gap in 2d semiconductors. 
For instance, for the BHZ - Hamiltonian adopted to HgTe quantum wells  $\Delta^{}_0\sim 10$meV~[\onlinecite{Zhang2008}].

We emphasize that even though we frequently use expressions like 'spin and its projections', we do not always mean the physical spin of the electron. 
For graphene or other systems which host Weyl- or Dirac-fermions it is better to think of the valley or sublattice degrees of freedom, rather than the electron spin. 
It is mainly due to the established tradition that we use this vocabulary. 

In strong magnetic field we replace the usual derivatives by the covariant ones $\displaystyle \partial^{}_\mu \to \partial^{}_\mu + i A^{}_\mu$, 
with the vector potential $A$ related to the magnetic field via $\displaystyle \nabla\times A = B$.
This condition can be realized by a number of gauges. We will use the central gauge 
\begin{equation}
A =  \frac{B}{2}
\left(
\begin{array}{c}
-y \\
 x\\
 0
\end{array}
\right),
\end{equation}
the choice which makes analytical calculations particularly convenient. 
Introducing complex coordinates $\displaystyle  z = x + iy$, $\displaystyle \bar z = x - iy$, and corresponding derivatives 
$\displaystyle\partial^{}_z = (\partial^{}_x - i\partial^{}_y)/2$, $\displaystyle \partial^{}_{\bar z} = (\partial^{}_x + i \partial^{}_y)/2$,
with the properties $\displaystyle\partial^{}_zz = \partial^{}_{\bar z}\bar z = 1,\;\; \partial^{}_z\bar z = \partial^{}_{\bar z}z = 0,$ we get 
$\displaystyle \nabla^{}_{-} \to 2\partial^{}_z + k^2\bar z    = A$, $\nabla^{}_{+} \to  2\partial^{}_{\bar z} - k^2 z = A^\dag$, and 
$\displaystyle \nabla^{}_{+}\nabla^{}_{-} \to (2\partial^{}_{\bar z} - k^2 z)( 2\partial^{}_z + k^2\bar z ) - 2k = A^\dag A - 2k$, where
\begin{equation}
\label{eq:MagnL}
k^2 = \frac{eB}{2\hbar} = \frac{1}{\ell^2}.
\end{equation}
$\ell = 1/k$ is referred to as the magnetic length. The operator $A$ annihilates the functions 
\begin{equation}
 \varphi^{}_{n} (r) = \frac{k}{\sqrt{\pi}}\frac{(k\bar z)^n}{\sqrt{n!}}e^{ -\frac{k^2}{2}z\bar z}
\end{equation}
i.e. $\displaystyle A \varphi^{}_{n} (r)  = 0$, for every positive integer $n$.
Hence, $\varphi^{}_{n} (r)$ is the lowest Landau level eigenfunction of the conventional operator of kinetic energy $\propto A^\dag A$. 
The Hilbert space of the lowest Landau level is infinitely degenerate, i.e. $n$ can assume every positive integer value between zero and infinity. 

\begin{figure}[t]
 \includegraphics[width=7cm]{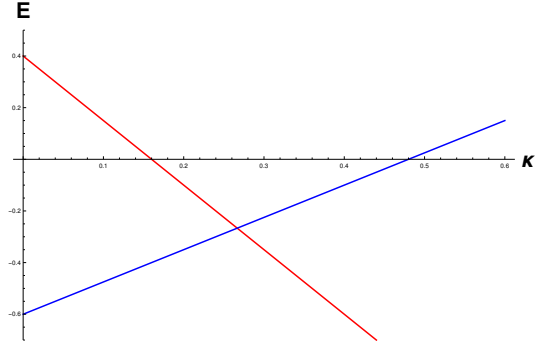}
\caption{
The energy spectrum of the Hamiltonian Eq.~(\ref{eq:BHZH}) in the lowest Landau level as function of the magnetic field. 
Red line shows $E^{}_-$ and the blue line $E^{}_+$ from Eq.~(\ref{eq:EVls}). 
In order to emphasize the split between the Landau zero points, the model parameters are chosen as $(2T^{}_-)^{-1}=-0.4$, $(2T^{}_+)^{-1}=0.8$, $\epsilon^{}_0=0.1$, $\Delta^{}_0=0.5$.
}
\label{fig:Fig1}
\end{figure}

In this notation, the Hamiltonian becomes 
\begin{equation}
 H = \Delta\Sigma^{}_{03} - \epsilon^{}_0\Sigma^{}_{00} + 2k\left({\cal C}^{}_+ T^{}_- + {\cal C}^{}_- T^{}_+\right) - 
 \left({\cal C}^{}_+ T^{}_- + {\cal C}^{}_- T^{}_+\right)A^\dag A  - iv\left({\cal D}^{}_+A + {\cal D}^{}_-A^\dag\right).
\end{equation}
The ground state (i.e. the eigenstate in the lowest Landau level) suffices the condition 
\begin{equation}
 \left[\left({\cal C}^{}_+ T^{}_- + {\cal C}^{}_- T^{}_+\right)A^\dag A  + iv\left({\cal D}^{}_+A + {\cal D}^{}_-A^\dag\right)\right]\psi = 0,
\end{equation}
which suggests two solutions:
\begin{equation}
\label{eq:EVs}
\psi^{}_{+,n}(r)  = \varphi^{}_n(r)\left(
\begin{array}{c}
0 \\ 1 \\ 0 \\  0
\end{array}
\right),
\;\;\; 
\psi^{}_{-,n}(r)  = \varphi^{}_n(r)\left(
\begin{array}{c}
0 \\ 0 \\ 1 \\  0
\end{array}
\right),
\end{equation}
which are unique up to a global phase and correspond to the two spin polarizations. 
The respective eigenvalues of the Hamiltonian for each spin projection in the lowest Landau level are found from the stationary Schr\"odinger equation
\begin{equation}
H\psi = E \psi,
\end{equation}
which yields for both spectral branches~[\onlinecite{Zhang2008}]
\begin{eqnarray}
\label{eq:EVls}
E^{}_+ = 2{T^{}_+}{k} - \epsilon^{}_0-\Delta^{}_0 &,& E^{}_- = 2{T^{}_-}{k} - \epsilon^{}_0+\Delta^{}_0  .
\end{eqnarray}
The existence of zero energy points is due to model's intrinsic Berry curvature~[\onlinecite{Zhang2006}].
In general, both eigenvalues may become zero at different magnetic fields:
\begin{eqnarray}
 E^{}_\pm = 0 &\rm for &  k^{}_\pm = \frac{\epsilon^{}_0 \pm \Delta^{}_0}{2T^{}_\pm}.
\end{eqnarray}
Below we refer to these points as the Landau zero points or modes. 
In the generic case, there is a split between zero energy points as shown in Fig.~\ref{fig:Fig1}.
For the material system of HgTe the Landau zero point is degenerate, i.e. $E^{}_+ = E^{}_-$ and the corresponding critical field is roughly $B\sim6$T~[\onlinecite{Zhang2008}].

The distance between both critical fields on the field axis is given by
\begin{equation}
\Delta k^{}_0 = |  k^{}_+ -  k^{}_- | = \left|\frac{\epsilon^{}_0}{2T^{}_+} - \frac{\epsilon^{}_0}{2T^{}_-}\right| + \left|\frac{\Delta^{}_0}{2T^{}_+} + \frac{\Delta^{}_0}{2T^{}_-}\right|.
\end{equation}

This model is also relevant for the case of graphene~[\onlinecite{Andrei2007,Novoselov2005,Kim2007}]. 
The Hamiltonian describing the chemically neutral gapless graphene in the strong magnetic field follows from Eq.~(\ref{eq:BHZH}) by putting the diagonal elements to zero.
Therefore, the eigenstates Eq.~(\ref{eq:EVs}) are eigenstates of the graphene Hamiltonian too, i.e. the whole analysis is applicable to graphene as well.
The only substantial difference is that for the gapless ($\Delta^{}_0=0$) and chemically neutral ($\epsilon^{}_0=0$) graphene, the energy of the the lowest Landau level is zero~[\onlinecite{Andrei2007}].

\section{Single-particle propagator in the lowest Landau level} 
\label{Sec:LLLGF}

The advanced (+) or retarded ($-$) Green's function in the lowest Landau level can be calculated using the spectral representation 
\begin{eqnarray}
\label{eq:BGF}
G^{\pm}_{r,r^\prime} &\sim& \sum^{\infty}_{n=0}\varphi^{}_n(r)\bar\varphi^{}_n(r')
\sum_{s=\pm}\frac{{\cal P}^{}_s}{E - E^{}_{s} \pm 0^+} ,
\end{eqnarray}
where $E^{}_s$ are the eigenvalues of the Hamiltonian for each spin projection in the lowest Landau level Eq.~(\ref{eq:EVls}) and the normalization will be fixed later. 
The projectors ${\cal P}^{}_{\pm}$ on the spin space 
\begin{equation}
{\cal P}^{}_+ = \left(
\begin{array}{cccc}
  0  &  0  &  0  & 0 \\
  0  &  1  &  0  & 0 \\
  0  &  0  &  0  & 0 \\
  0  &  0  &  0  & 0 
\end{array}
\right) 
%= \frac{1}{2}\left[\Sigma^{}_{00} - \Sigma^{}_{03}\right] 
\;\;\;  {\rm and} \;\;\;
{\cal P}^{}_- = \left(
\begin{array}{cccc}
  0  &  0  &  0  & 0 \\
  0  &  0  &  0  & 0 \\
  0  &  0  &  1  & 0 \\
  0  &  0  &  0  & 0 
\end{array}
\right)
\end{equation}
are idempotent and orthogonal matrices with properties
$\displaystyle {\cal P}^{}_{+}{\cal P}^{}_{-} = 0,\;\; {\cal P}^{}_{+}{\cal P}^{}_{+} = {\cal P}^{}_{+},\;\;\;  {\cal P}^{}_{-}{\cal P}^{}_{-} = {\cal P}^{}_{-}$.
The summation over all $n$ yields 
\begin{eqnarray}
\sum^{\infty}_{n=0}\varphi^{}_n(r)\bar\varphi^{}_n(r') &=& \frac{k^2}{\pi} e^{-\frac{k^2}{2}(|z|^2+|z^\prime|^2)} \sum_{n=0}^\infty \frac{(k^2\bar zz^\prime)^n}{n!} 
= \frac{k^2}{\pi} e^{-\frac{k^2}{2}(|z|^2+|z^\prime|^2 - 2\bar zz^\prime)},
\end{eqnarray}
which then gives for the Green's function [\onlinecite{Wegner1983,Chakravarty2007}]
\begin{equation}
\label{eq:GFLLL}
G^{\pm}_{r,r^\prime} (E)=  \frac{k^2}{2\pi} e^{-\frac{k^2}{2}(|z|^2+|z^\prime|^2 - 2\bar zz^\prime)} 
\sum_{s=\pm}\frac{{\cal P}^{}_s}{E - E^{}_{s} \pm 0^+}.
\end{equation}
Notably, the local Green's function ($r=r'$) is a coordinate independent constant. 
The propagator is normalized this way in order to satisfy the usual sum rule 
\begin{equation}
\label{eq:SumRule}
\mp\intop^\infty_{-\infty}\frac{dE}{\pi}~ {\rm Im}~{\rm tr} G^{\pm}_{r,r} (E) = \frac{k^2}{\pi} = \frac{eB}{h}, 
\end{equation}
where the trace operator acts only on the spin space. Eq.~(\ref{eq:SumRule}) gives the number of the elementary flux quanta $\phi^{}_0=h/e$ per unit volume.
In the real time representation, the Green's function represents a simple collection of undamped harmonic functions with the period determined by the 
eigenenergies of the lowest Landau level modes 
\begin{equation}
\label{eq:GFLLLt}
G^{\pm}_{r,r^\prime}(t) = \mp i\frac{k^2}{2\pi} e^{-\frac{k^2}{2}(|z|^2+|z^\prime|^2 - 2\bar zz^\prime)}\sum_{s=\pm}{\cal P}^{}_s e^{\pm iE^{}_st},
\end{equation}
the initial time is assumed to be at zero. The Green's function is totally separable on the space-time. 

For the case of chemically neutral gapless graphene the Green's function becomes particularly simple~[\onlinecite{Chakravarty2007}]:
\begin{equation}
\label{eq:GFgraph}
 G^{\pm }_{rr'} (E) = \frac{k^2}{2\pi}\frac{1}{E\pm i0^+} e^{-\frac{k^2}{2}(|z|^2+|z^\prime|^2 - 2\bar zz^\prime)}  [{\cal P}^{}_+ + {\cal P}^{}_-],
\end{equation}
i.e. in the real time representation it is just a step function $\theta(t)$.

\section{Dressing of the single-particle propagator due to the disorder}
\label{Sec:Dress}

The disorder is introduced in the form of the fluctuating chemical potential $v(r)$, which couples in the spin space to the unity matrix $\Sigma^{}_{00}$, 
with the white noise correlator:
\begin{equation}
\label{eq:WNC}
\langle v^{}_r \rangle^{}_g =0,\;\; \langle v^{}_{r^{}_1} v^{}_{r^{}_2}\rangle^{}_g = g\delta^{}_{r^{}_1 r^{}_2}.
\end{equation}
The averaged propagator reads 
\begin{equation}
\label{eq:GFfull}
\bar G^{\pm}_{r^{}_1 r^{}_2} = \langle [(G^\pm)^{-1} + v\Sigma^{}_{00}]^{-1}_{r^{}_1 r^{}_2} \rangle^{}_g .
\end{equation}
To perform the disorder average perturbatively we expand the propagator in powers of $v$: 
\begin{eqnarray}
\nn
&\displaystyle
\bar G^{\pm}_{r^{}_1 r^{}_2} =  \langle G^{\pm}_{r^{}_1 r^{}_2} 
-G^{\pm}_{r^{}_1 x^{}_1}v^{}_{x^{}_1}G^{\pm}_{x^{}_1 r^{}_2}
+G^{\pm}_{r^{}_1 x^{}_1}v^{}_{x^{}_1}G^{\pm}_{x^{}_1 x^{}_2}v^{}_{x^{}_2}G^{\pm}_{x^{}_2 r^{}_2}
-G^{\pm}_{r^{}_1 x^{}_1}v^{}_{x^{}_1}G^{\pm}_{x^{}_1 x^{}_2}v^{}_{x^{}_2}G^{\pm}_{x^{}_2 x^{}_3}v^{}_{x^{}_3}G^{\pm}_{x^{}_3 r^{}_2} +&
\\
&\displaystyle 
 G^{\pm}_{r^{}_1 x^{}_1}v^{}_{x^{}_1}G^{\pm}_{x^{}_1 x^{}_2}v^{}_{x^{}_2}G^{}_{x^{}_2 x^{}_3}v^{}_{x^{}_3}G^{\pm}_{x^{}_3 x^{}_4}v^{}_{x^{}_4}G^{\pm}_{x^{}_4 r^{}_2}
-G^{\pm}_{r^{}_1 x^{}_1}v^{}_{x^{}_1}G^{\pm}_{x^{}_1 x^{}_2}v^{}_{x^{}_2}G^{}_{x^{}_2 x^{}_3}v^{}_{x^{}_3}G^{\pm}_{x^{}_3 x^{}_4}v^{}_{x^{}_4}G^{\pm}_{x^{}_4 x^{}_5}v^{}_{x^{}_5}G^{\pm}_{x^{}_5 r^{}_2} \cdots
\rangle^{}_g ,&
\hspace{6mm}
\end{eqnarray}
where the summation over multiple indices is understood. Because of Eq.~(\ref{eq:WNC}) all terms with an odd number of potentials $v$ vanish. The series then becomes 
\begin{eqnarray}
\nn
&\displaystyle
\bar G^{\pm}_{r^{}_1 r^{}_2} = \langle G^{\pm}_{r^{}_1 r^{}_2} + 
G^{\pm}_{r^{}_1 x^{}_1}v^{}_{x^{}_1}G^{\pm}_{x^{}_1 x^{}_2}v^{}_{x^{}_2}G^{\pm}_{x^{}_2 r^{}_2} +
G^{\pm}_{r^{}_1 x^{}_1}v^{}_{x^{}_1}G^{\pm}_{x^{}_1 x^{}_2}v^{}_{x^{}_2}G^{\pm}_{x^{}_2 x^{}_3}v^{}_{x^{}_3}G^{\pm}_{x^{}_3 x^{}_4}v^{}_{x^{}_4}G^{\pm}_{x^{}_4 r^{}_2}
&\\
\label{eq:PT}
&\displaystyle+
G^{\pm}_{r^{}_1 x^{}_1}v^{}_{x^{}_1}G^{\pm}_{x^{}_1 x^{}_2}v^{}_{x^{}_2}G^{\pm}_{x^{}_2 x^{}_3}v^{}_{x^{}_3}G^{\pm}_{x^{}_3 x^{}_4}v^{}_{x^{}_4}G^{\pm}_{x^{}_4 x^{}_5}v^{}_{x^{}_5}G^{\pm}_{x^{}_5 x^{}_6}v^{}_{x^{}_6}G^{\pm}_{x^{}_6 r^{}_2}
\cdots
\rangle^{}_g .&
\end{eqnarray}

The Green's function Eq.~(\ref{eq:GFLLL}) is spanned by the spin projectors ${\cal P}^{}_s$. Therefore, only the disorder diagonal in the spin space is of importance. 
Besides the randomly fluctuating chemical potential considered here, these might include the randomly fluctuating gap, which couples to $\Sigma^{}_{03}$, the random "chiral" chemical potential 
($\Sigma^{}_{30}$), or the random "chiral" mass ($\Sigma^{}_{33}$). 
Each product of these matrices with ${\cal P}^{}_s$ projects them bar the sign back onto ${\cal P}^{}_s$ again. 
Therefore, the perturbative series Eq.~(\ref{eq:PT}) does not depend on a particular disorder type and our analysis is generic and disorder type independent.

\begin{figure}[t]
\includegraphics[width=7cm]{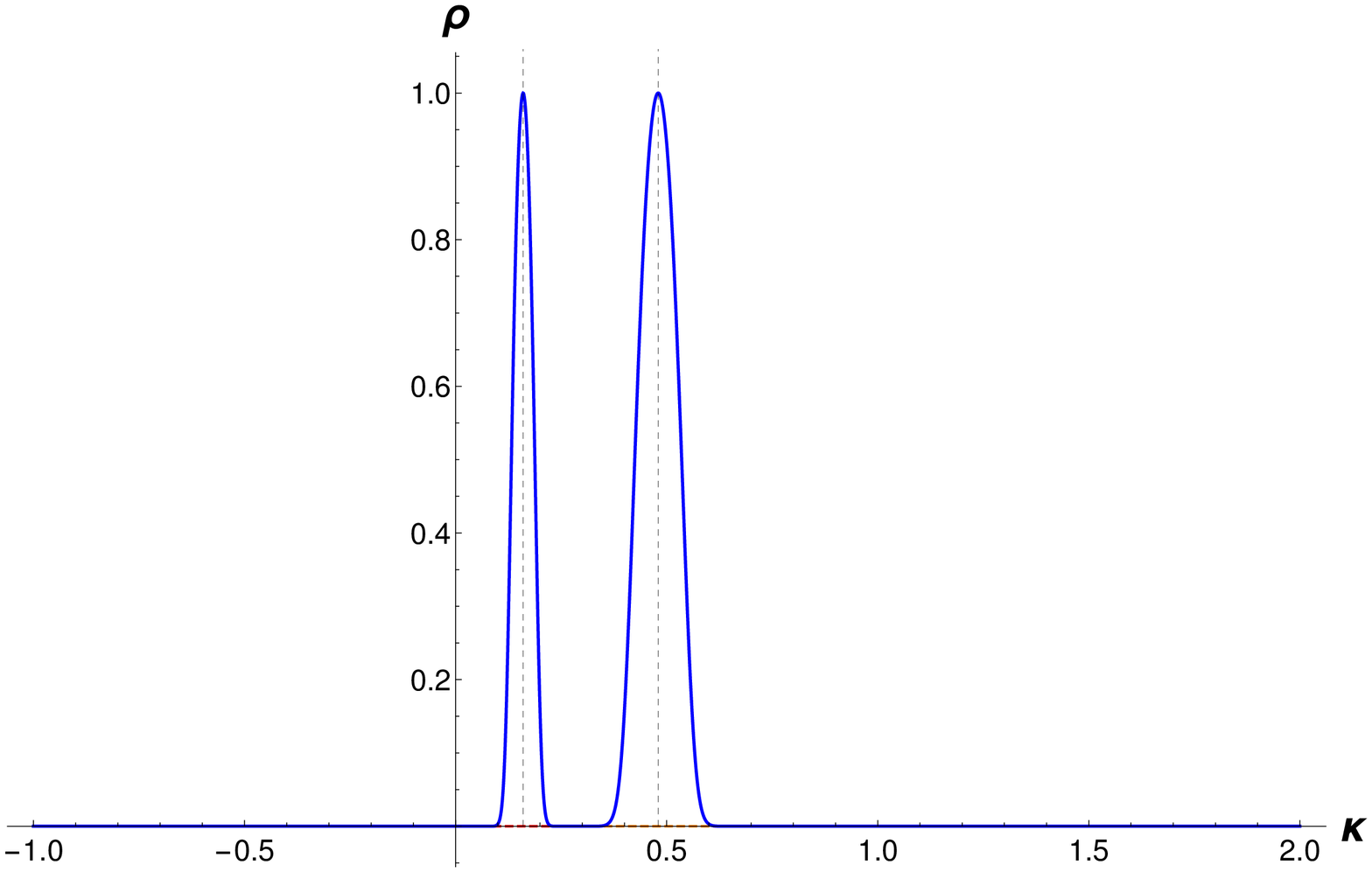}
\hspace{5mm}
\includegraphics[width=7cm]{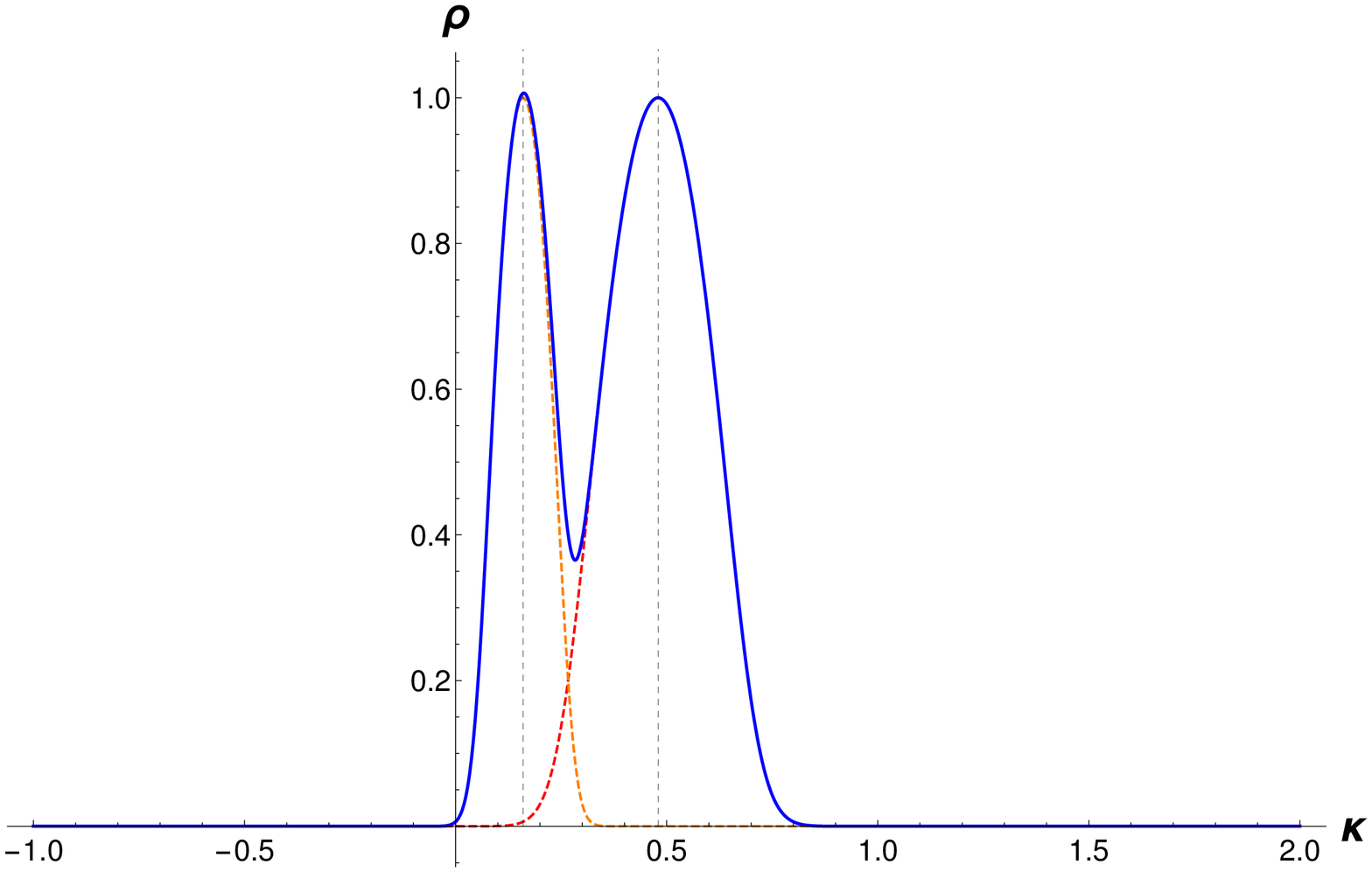}

\includegraphics[width=7cm]{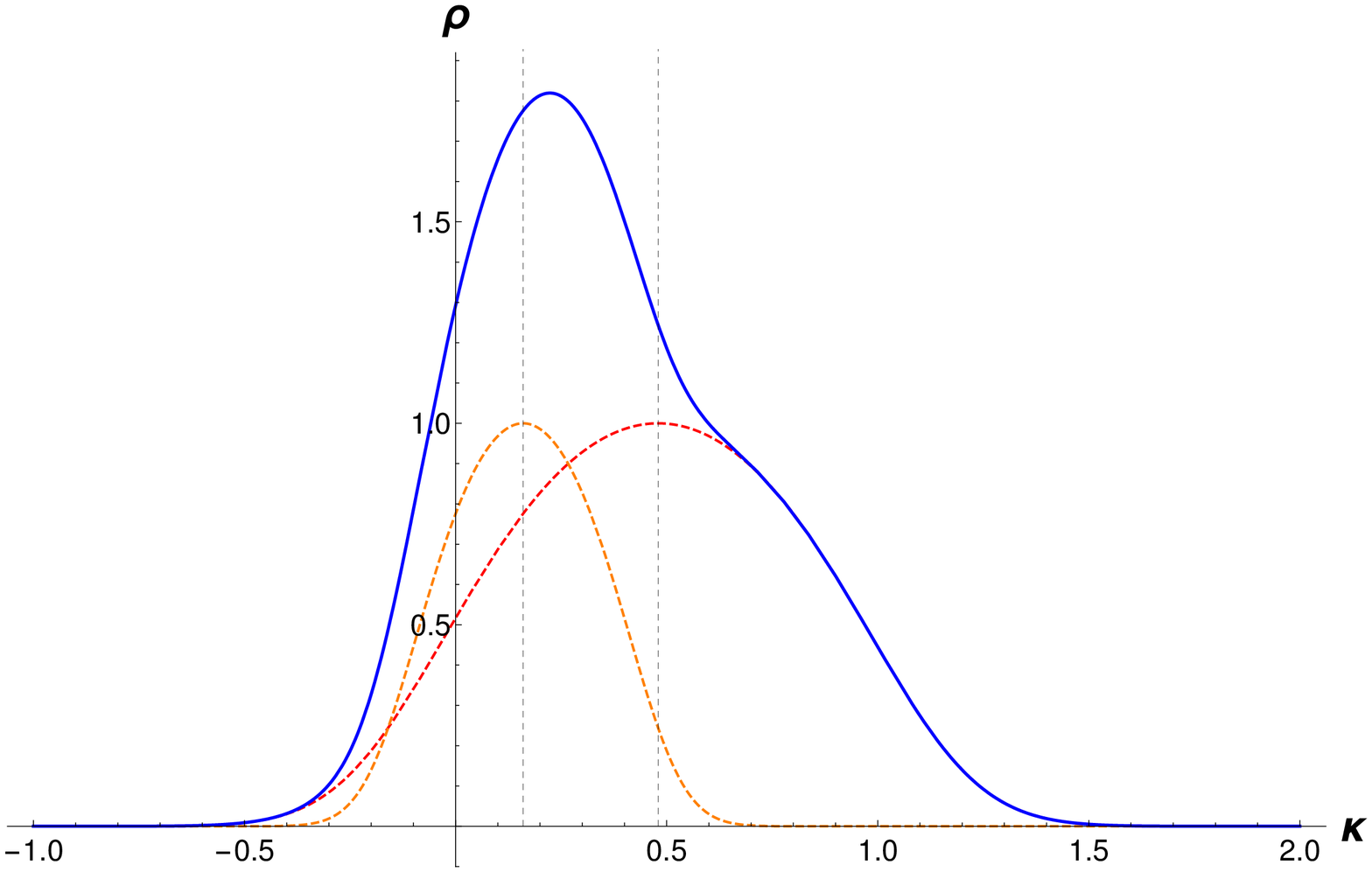}
\hspace{5mm}
\includegraphics[width=7cm]{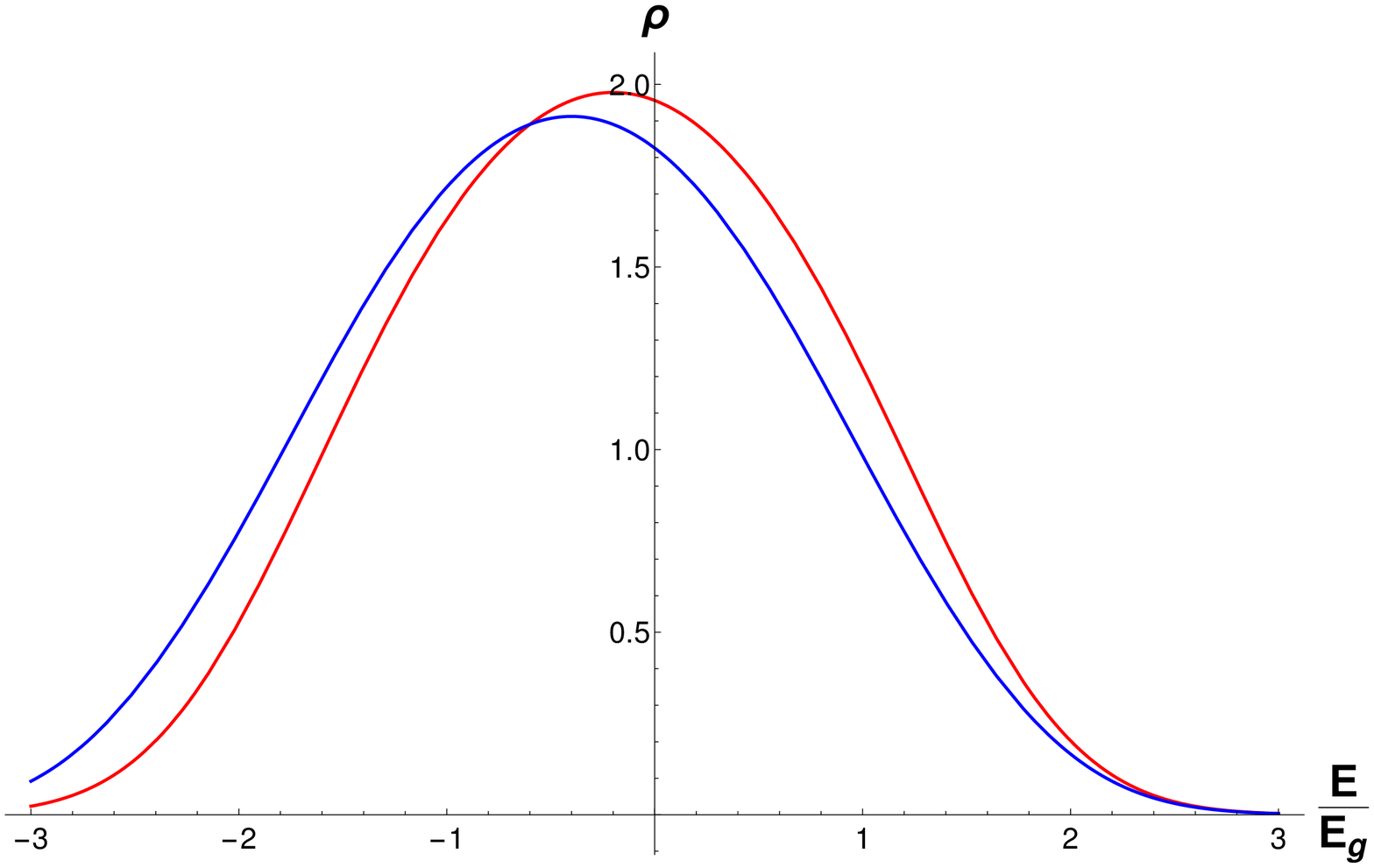}
\caption{
Density of states (DOS) Eq.~(\ref{eq:DOS}) calculated for the spectrum shown in Fig.~\ref{fig:Fig1} in units of $\frac{1}{\pi^{5/2}}\frac{k^2}{E^{}_g}$.
Top and bottom left: 
Evolution of the DOS in the band center ($E=0$)  with increasing disorder strength as function of the magnetic field $k\sim\sqrt{B}$.
Dashed lines emphasize the contributions to the net density of states from each Landau zero mode. The DOS is centered around the Landau zero points.
Multiplying this function by the factor given in Eq.~(\ref{eq:CondLZ}) gives the dc conductivity according to the Einstein relation Eq.~(\ref{eq:EinsteinRel}). 
Bottom right: DOS as function of the dimensionless energy $E/E^{}_g$ for the magnetic fields adjusted to each of the Landau zero points in 
Fig.~\ref{fig:Fig1}. Here, the color scheme of Fig.~\ref{fig:Fig1} is preserved.
}
\label{fig:Fig2}
\end{figure}

The exact Green's function of disordered electrons in the lowest Landau level was obtained by Wegner in Ref.~[\onlinecite{Wegner1983}]. 
The constraint condition Eq.~(\ref{eq:SumRule}) changes it to, cf. Appendix~\ref{app:WegnerGF}.
\begin{equation}
\label{eq:WegnerF}
\bar G^{\pm}_{rr'} (E) =  
\frac{k^2}{\pi}e^{-\frac{k^2}{2}(|r|^2+|r'|^2-2\bar r r')}\sum_{s=\pm} {\cal F}^{\pm}_s(E){\cal P}^{}_s, 
\end{equation}
where the frequency dependent part
\begin{equation}
{\cal F}^{\pm}_s(E) = \eta^{}_{s}(E) \mp i \rho^{}_s(E),
\end{equation}
has the following explicit expressions for the real 
\begin{equation}
\label{eq:WRE}
\eta^{}_{s}(E) = \frac{1}{E^{}_g}\left(
\frac{2}{\pi}\frac{\displaystyle e^{\nu^2_s}\int^{\nu^{}_s}_0dt~ e^{t^2}}{\displaystyle 1+\left(\frac{2}{\sqrt{\pi}}\int^{\nu^{}_s}_0dt~ e^{t^2}\right)^2} - \nu^{}_s
\right), 
\end{equation}
and imaginary part~[\onlinecite{Brezin1984,Hikami1984a}]
\begin{equation}
\label{eq:WIM}
\rho^{}_{s}(E) = \frac{1}{\sqrt{\pi} E^{}_g}\frac{\displaystyle e^{\nu^2_s}}{\displaystyle 1+\left(\frac{2}{\sqrt{\pi}}\int^{\nu^{}_s}_0dt~ e^{t^2}\right)^2}.
\end{equation}
They depend on the dimensionless energy 
\begin{equation}
\nu^{}_s = \frac{E-E^{}_s}{E^{}_g},\;\;\, {\rm where}\;\; E^2_g=\frac{gk^2}{4\pi}. 
\end{equation}
In the chosen units the disorder related energy $E^{}_g$ is a dimensionless quantity. It represents essentially the ratio of two relevant lengths $E^{}_g\sim l^{}_\lambda/\ell$: 
the magnetic length $\ell\sim1/k$ and the disorder related length $l^{}_\lambda\sim\sqrt{g}$. 
The collective density of states 
\begin{equation}
\label{eq:DOS}
\rho(E) =\mp\frac{1}{\pi} {\rm Im}~{\rm tr} G^{\pm}_{rr}(E) = \frac{1}{\pi^{5/2}}\frac{k^2}{E^{}_g}
\sum_{s=\pm}\frac{\displaystyle e^{\nu^2_s}}{\displaystyle 1+\left(\frac{2}{\sqrt{\pi}}\int^{\nu^{}_s}_0dt~ e^{t^2}\right)^2}.
\end{equation}
is correctly normalized in accordance with Eq.~(\ref{eq:SumRule}). 
Fig.~\ref{fig:Fig2} shows Eq.~(\ref{eq:DOS}) for spectrum from Fig.~\ref{fig:Fig1}. 
For weak disorder strength the density of states appears in the form of two sharp peaks placed symmetrically around the Landau zeros. 
It is plotted in units of $\frac{1}{\pi^{5/2}}\frac{k^2}{E^{}_g}\sim(\ell l^{}_\lambda)^{-1}$, ${\ell l^{}_\lambda}$ being the parametric volume constructed from the two specific lengths of the model.
The peaks become boarder with increasing disorder strength and overlap with each other until they merge to a single structure. 

Performing the Fourier transformation we obtain the Wegner's propagator in the time  representation. 
It comprises two parts, one periodically oscillating in time and the second localized in the time with the maximum at zero time, cf. Appendix~\ref{app:WegnerGF}:
\begin{equation}
\label{eq:WGFT}
 {\cal F}^{\pm}_s(t) = \mp\frac{i}{2}e^{\pm iE^{}_s t}\Omega(E^{}_gt).
\end{equation}
The real part of the Wegner's propagator is antisymmetric in time, while the imaginary part is symmetric. 
Notably, the period of oscillations is still determined only by energies of the clean system, while the temporarily localized part $\Omega(E^{}_gt)$ depends only on the disorder related energy $E^{}_g$. It is shown in Fig.~\ref{fig:Fig3} and represents a smooth and strongly damped oscillating function symmetrically placed around $t=0$. The series of $\Omega(E^{}_gt)$ contains only 
even powers of $E^{}_gt$ and goes for small times as $\Omega(E^{}_gt)\sim 1-{\rm const}\cdot t^2$.

\section{Mean squared displacement of the disordered system }
\label{Sec:OptCon}

The access to the diffusion goes via the mean squared displacement
\begin{equation}
\label{eq:MSqR}
\langle r^2_{\mu}(t)\rangle = \frac{\displaystyle{\rm tr}\sum_r r^2_\mu P^{}_{r0}(t)}{\displaystyle{\rm tr}\sum_r P^{}_{r0}(t)}, 
\end{equation}
where $r^{}_\mu$ is the position operator and $ P^{}_{rr'}(t) $ is the return probability density defined as
\begin{equation}
 P^{}_{rr'}(t) = \int\frac{dE}{2\pi}~e^{-iEt}  P^{}_{rr'}(E) ,
\end{equation}
where 
\begin{equation}
\label{eq:da2pgf}
 P^{}_{rr'}(E) = \langle G^+_{rr'}(E)G^-_{r'r}(E)\rangle^{}_g,
\end{equation}
is the disorder averaged two-particles propagator. The large time asymptotics of the mean squared displacement is expected to be of the form~[\onlinecite{Goldenfeld,Huang,Chaikin}] 
\begin{equation}
\label{eq:MsqD}
\lim_{t\to\infty} \langle r^2_{\mu}(t)\rangle\to 2Dt^{1+\alpha} + {\rm const},
\end{equation}
where $D$ is the diffusion coefficient and the exponent $\alpha$ is referred to as the anomalous dimension and can be either positive (superdiffusion) or negative 
(subdiffusion)~[\onlinecite{Goldenfeld}]. For instance, for the ordinary diffusion $\alpha=0$ with $P_{r0}\sim\exp[r^2/Dt]/Dt$, which can be easily verified by evaluating Eq.~(\ref{eq:MSqR})~[\onlinecite{Ziegler2012}]. 
In the linear regime, the relation between the mean squared displacement and diffusion is established via 
\begin{equation}
\label{eq:MsqD1}
\left.\frac{d}{dt}\right|_{t=0} \langle r^2_{\mu}(t)\rangle =  2D.
\end{equation}

To start with, we evaluate Eq.~(\ref{eq:MSqR}) for the case of clean system. Because the real time representation of the Green's function of clean system Eq.~(\ref{eq:GFLLLt}) 
is totally separable into the spatial and temporal parts, the same is also valid for the return probability density $P^{}_{r,r'}(t) = R^{}({r,r'})T(t)$. 
Hence, the time dependent parts cancel each other in Eq.~(\ref{eq:MSqR}) exactly and we are left with a simple result
\begin{equation}
\label{eq:msdcs}
\langle r^2_{\mu}(t)\rangle = \frac{\ell^2}{2}, 
\end{equation}
where we use explicitly the magnetic length defined in Eq.~(\ref{eq:MagnL}) and the factor $1/2$ is attributed to the angular average. 
The result is transparent: it gives the expected position of an electron to be a circle with radius $\ell$ around the position of the flux tube 
piercing the sample at the origin of coordinates. 

Our task is to determine the diffusion coefficient of the disordered system through the direct evaluation of Eq.~(\ref{eq:MSqR}), in order to compute the conductivity from the Einstein relation
\begin{equation}
\label{eq:EinsteinRel} 
\sigma = \frac{e^2}{\hbar} D\rho(E),
\end{equation}
where $\rho(E)$ is the density of states discussed in the previous paragraph.

\begin{figure}[t]
\includegraphics[width=5cm]{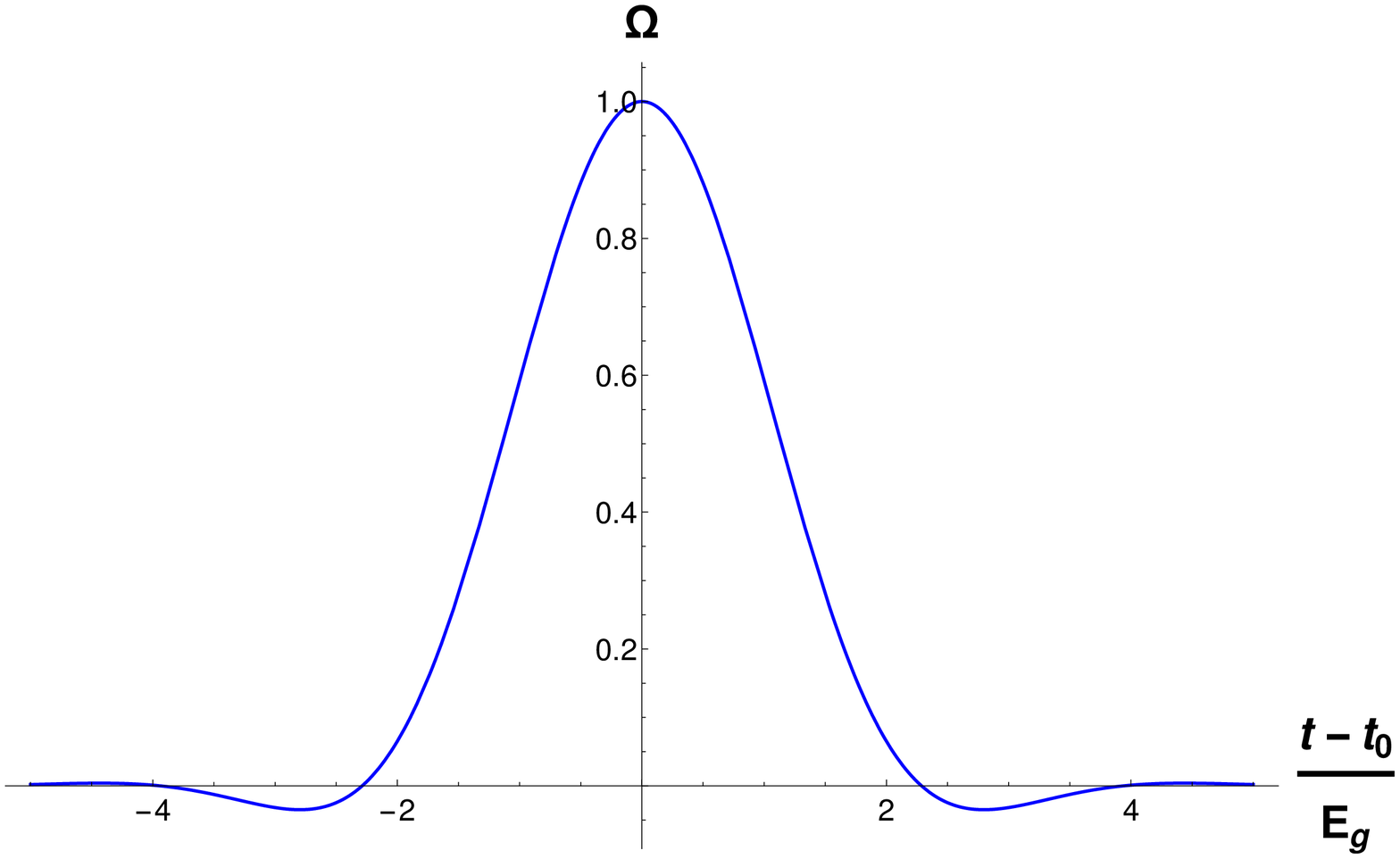}
\hspace{2mm}
\includegraphics[width=5cm]{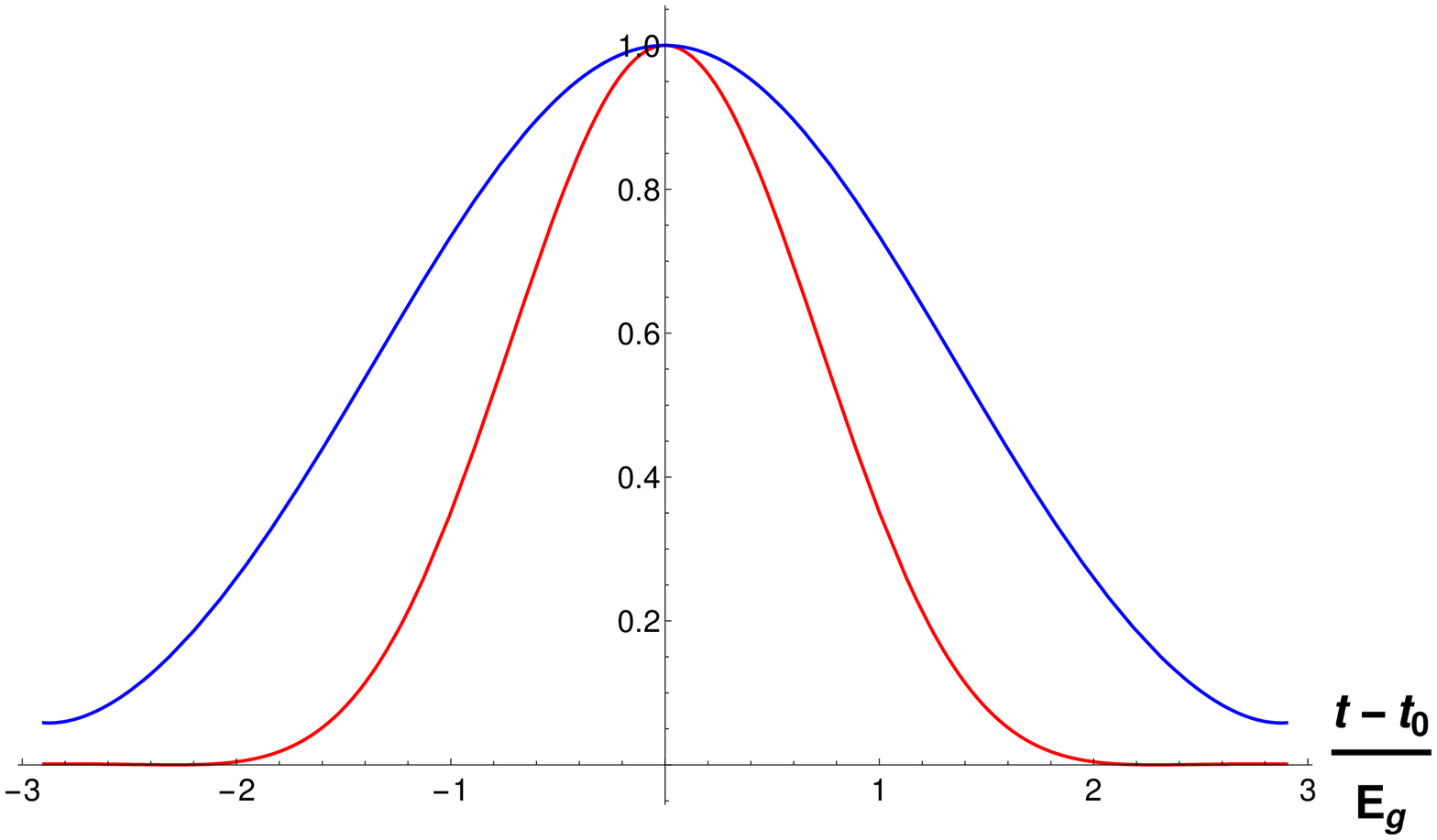}
\hspace{2mm}
\includegraphics[width=5cm]{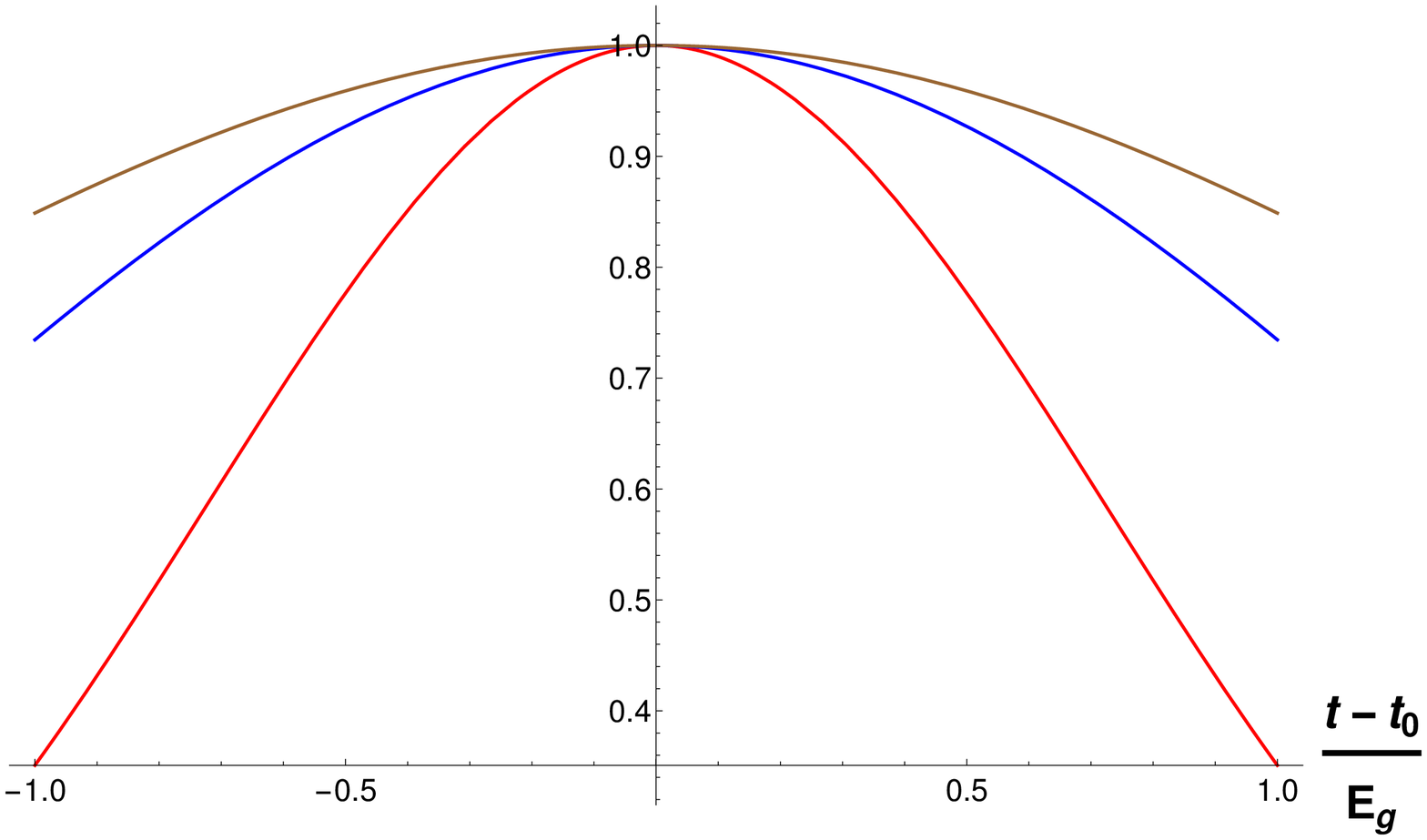}
\caption{
Left: The shape of the function $\Omega(E^{}_gt)$ from Eq.~(\ref{eq:WGFFT}) determined to the order $(E^{}_gt)^{60}$.
Middle: The second time convolution of the Wegner's function Eq.~(\ref{eq:SecCon}) (blue line) in comparison to $\Omega^2(E^{}_gt)$ (red line). 
Right: Same distributions as in the middle panel plus the fourth time convolution (brown line) Eq.~(\ref{eq:ForCon}) for small times.
All curves are normalized to 1 at their maxima in respective units. The higher is the order of the convolution, the broader is its distribution.
}
\label{fig:Fig3}
\end{figure}

A rigorous evaluation of the full perturbative series for the two-particles propagator $\langle G^+_{r,0}G^-_{0,r}\rangle^{}_g$ along the lines of Wegner's calculations
for the single-particle propagator is principally impossible. 
The reason for this is the spatial decoherence in higher orders of perturbative expansion, cf. Appendix~\ref{app:teciss}:  
While the disorder does not affect the spatial dependence of the single-particle propagator, it does so for the two-particles propagator. 
Thus, we need to consider the spatial averages. 
We evaluate both expressions from the numerator and denominator of Eq.~(\ref{eq:MSqR}) perturbatively. Evaluation of all perturbative diagrams to order $g^3$ yields for the numerator of Eq.~(\ref{eq:MSqR}) 
\begin{eqnarray}
\nn
&\displaystyle
{\rm tr}\sum_r r^2_\mu P^{}_{r0}(E) = \frac{1}{4\pi}\frac{1}{E^2_g}\sum^{}_{s=\pm} (2X^{}_s)^2\left[\frac{1}{2} + (2X^{}_s)^2 + 2(2X^{}_s)^4 + \frac{167}{36}(2X^{}_s)^6 + \cdots \right. 
&\\
\label{eq:WCnd2cc}
&\displaystyle + 
\left.\left(\frac{3}{4} (2X^{}_s)^4 + \frac{343}{72}(2X^{}_s)^6 + \cdots \right)\cos2\phi^{}_s + 
\left(\frac{139}{72}(2X^{}_s)^6 + \cdots \right)\cos4\phi^{}_s  + \cdots \right],
&
\end{eqnarray}
where $X^2_s(E) = E^2_g [\eta^2_s(E) + \rho^2_s(E)]$ and $\phi^{}_s(E) = {\rm arctan}\left[\frac{\rho^{}_s(E)}{\eta^{}_s(E)}\right]$. 
According to Eqs.~(\ref{eq:WRE}) and (\ref{eq:WIM}), $X^2_s(E)$ and $\phi^{}_s(E)$ are dimensionless functions of the argument $\nu^{}_s=(E-E^{}_s)/E^{}_g$.
The analogous computation for the denominator of Eq.~(\ref{eq:MSqR}) yields 
\begin{eqnarray}
\nn
&\displaystyle
{\rm tr}\sum_r P^{}_{r0}(E) = \frac{k^2}{4\pi}\frac{1}{E^2_g}\sum^{}_{s=\pm} (2X^{}_s)^2\left[1 + (2X^{}_s)^2 + \frac{3}{2} (2X^{}_s)^4 + \frac{13}{4} (2X^{}_s)^6 + \cdots \right. 
&\\
\label{eq:WCnd2nn}
&\displaystyle + 
\left.\left((2X^{}_s)^4 + \frac{9}{2}(2X^{}_s)^6 + \cdots \right)\cos2\phi^{}_s + 
\left(\frac{5}{2}(2X^{}_s)^6 + \cdots \right)\cos4\phi^{}_s + \cdots \right].
&
\end{eqnarray}
Technical details of the evaluation and results for each individual diagram are summarized in Appendices~\ref{app:teciss} and~\ref{app:MSD}. 

At first one can try to Fourier transform each individual term in both series. Essentially, the Fourier transformed of Eqs.~(\ref{eq:WCnd2cc}) and~(\ref{eq:WCnd2nn}) are 
given in terms of even time convolutions of the temporal part of the Wegner's propagator Eq.~(\ref{eq:WGFT}) 
\begin{eqnarray}
\label{eq:SecCon}
\int^\infty_{-\infty}\frac{dE}{\pi} e^{iEt} X^2_s(E) &\sim& \int^\infty_{-\infty} d\tau~ \Omega(E^{}_g\tau)\Omega(E^{}_g[\tau+t]),  \\
\int^\infty_{-\infty}\frac{dE}{\pi} e^{iEt} X^4_s(E) &\sim& 
\int^\infty_{-\infty} d\tau^{}_1 \int^\infty_{-\infty} d\tau^{}_2 \int^\infty_{-\infty}d\tau^{}_3 ~ \Omega(E^{}_g\tau^{}_1)\Omega(E^{}_g[\tau^{}_1+\tau^{}_3])\\
\label{eq:ForCon}
&\times& \Omega(E^{}_g\tau^{}_2)\Omega(E^{}_g[\tau^{}_2+\tau^{}_3+t]),  
\end{eqnarray}
and so on. In the practice though, this can be done to an extend only. To the best of our knowledge, the function $\Omega(E^{}_g\tau)$ itself is not tabulated and its polynomial series has to be truncated at some order. For practical calculations we use the polynomial with 31 terms (order $(E^{}_gt)^{60}$), which is sufficient to determine the second convolution Eq.~(\ref{eq:SecCon}) only to the order $(E^{}_gt)^{20}$, and the forth convolution Eq.~(\ref{eq:ForCon}) only to the order $(E^{}_gt)^{6}$, cf. Fig.~\ref{fig:Fig3}. The convolutions have a typical bell-curve shape.
This result suffices to recognize that every higher order convolution is broader and therefore, the large time asymptotics of the mean squared displacement is determined 
by the higher order terms in the series Eqs.~(\ref{eq:WCnd2cc}) and (\ref{eq:WCnd2nn}). Another aspect is that since the Wegner's propagator is uniquely separable in temporal and spatial 
variables, the large time asymptotics of the numerator and denominator of Eq.~(\ref{eq:MSqR}) differ only in amplitudes but not in functional behavior. Hence, we can expect that for
large times the mean squared displacement approaches a time independent constant, similarly to the case of the clean system Eq.~(\ref{eq:msdcs}) and the time evolution ends up in a stationary state.
In this state all electrons are again localized at fixed orbits and no currents can flow through the sample. 
If there be a diffusion in the system, i.e. the regime linear in time with a positive slope, then it can only emerge due to the superposition of all convolutions at intermediate time scales.

\begin{figure}[t]
\includegraphics[width=9cm]{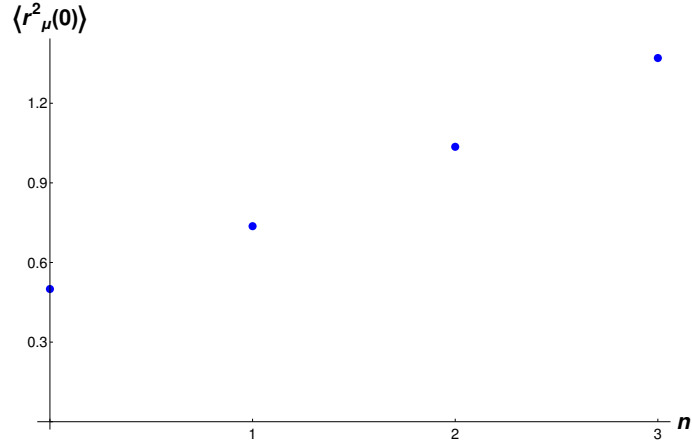}
\caption{
The initial mean squared displacement $\langle r^2_\mu(0)\rangle$ Eq.~(\ref{eq:InMSqD}) in units of $k^{-2}$ evaluated to different order ($n=0,1,2,3$) of perturbative expansion. 
}
\label{fig:Fig4}
\end{figure}

To give these general arguments a more firm footing we propose a resummation procedure leaned on the general idea of the renormalization group. 
We start with the considerations of the mean squared displacement at time zero 
\begin{equation}
\label{eq:MSqRE}
\langle r^2_{\mu}(0)\rangle = \frac{\displaystyle{\rm tr}\int^\infty_{-\infty} dE\sum_r r^2_\mu P^{}_{r0}(E)}{\displaystyle{\rm tr}\int^\infty_{-\infty} dE \sum_r P^{}_{r0}(E)}. 
\end{equation}
Changing to the integration over $\nu^{}_s$ we then have 
\begin{equation}
\label{eq:InMSqD}
k^2  \langle r^2_{\mu}(0)\rangle = \frac{\displaystyle \sum^{}_{s=\pm}~\int^\infty_{-\infty} d\nu^{}_s~f^{(2)}(\nu^{}_s)}{\displaystyle \sum^{}_{s=\pm}\int^\infty_{-\infty} d\nu^{}_s~f^{(0)}(\nu^{}_s)} = 
\frac{\langle f^{(2)}\rangle}{\langle f^{(0)}\rangle},
\end{equation}
where 
\begin{eqnarray}
\langle f^{(2)}\rangle = {\rm tr}\int^{\infty}_{-\infty}d\nu^{}_s\sum_{r,s=\pm} r^2_\mu P^{}_{r0}(\nu^{}_s) , & \; {\rm and} \; &
\langle f^{(0)}\rangle = {\rm tr}\int^{\infty}_{-\infty}d\nu^{}_s \sum_{r,s=\pm} P^{}_{r0}(\nu^{}_s).
\end{eqnarray}
Each of both spin projections contribute equally in $t=0$ limit and we can skip the summation over $s$.
In Fig.~\ref{fig:Fig4} we plot the right hand side of Eq.~(\ref{eq:InMSqD}) evaluated for the increasing order $n$ of perturbative expansion. 
 
With larger $n$ the statical mean squared displacement approaches a linear asymptotics as function of $n$. Among all possible diagrammatic channels, there is a unique channel which suffices 
this property. This includes all diagrams of the so-called ladder channel, cf. Appendix~\ref{app:MSD}. Assembling all ladder diagrams yields the asymptotics of the two-particles propagator 
in the form of an infinite series
\begin{equation}
P^{\rm lad}_{r0}(E) \approx  \frac{1}{4E^2_g}\left(\frac{k^2}{\pi}\right)^2\sum_{s=\pm} ~ \sum^\infty_{n=1}\frac{(2X^{}_s)^{2n}}{n}\exp\left[-\frac{k^2r^2}{n}\right].
\end{equation}
This expression gives rise to the series element of the denominator of Eq.~(\ref{eq:InMSqD}) in the form
\begin{equation}
\label{eq:LadDen}
\langle f^{(0)}_n\rangle^{\rm lad} = \frac{k^2}{4\pi E^{2}_g}\sum_{s=\pm}\int^\infty_{-\infty} d\nu~ (2X^{}_s)^{2n},
\end{equation}
and for the numerator 
\begin{equation}
\label{eq:LadNum}
\langle f^{(2)}_n\rangle^{\rm lad} = \frac{1}{4\pi E^{2}_g}\sum_{s=\pm}\int^\infty_{-\infty} d\nu~ \frac{n}{2} (2X^{}_s)^{2n}.
\end{equation}
Because the argument $2X^{}_s$ is larger than 1 at its maximum the series do not converge. Hence, at every finite expansion order, the behavior of both numerator and denominator of 
the mean squared displacement are dominated by the respective last terms in the series, which then cancel within the formula for $\langle r^2_{\mu}(0)\rangle$ Eq.~(\ref{eq:MSqRE}). 
In this way the linear dependence of $\langle r^2_{\mu}(0)\rangle\sim n/2$  visible in Fig.~\ref{fig:Fig4} emerges. 
All remaining channels exhibit either slower logarithmic grow or decay with increasing $n$, which allows to neglect them.

\section{Equation of motion for the mean squared displacement}
\label{Sec:EMMSD}

\begin{figure}[t]
\includegraphics[width=7cm]{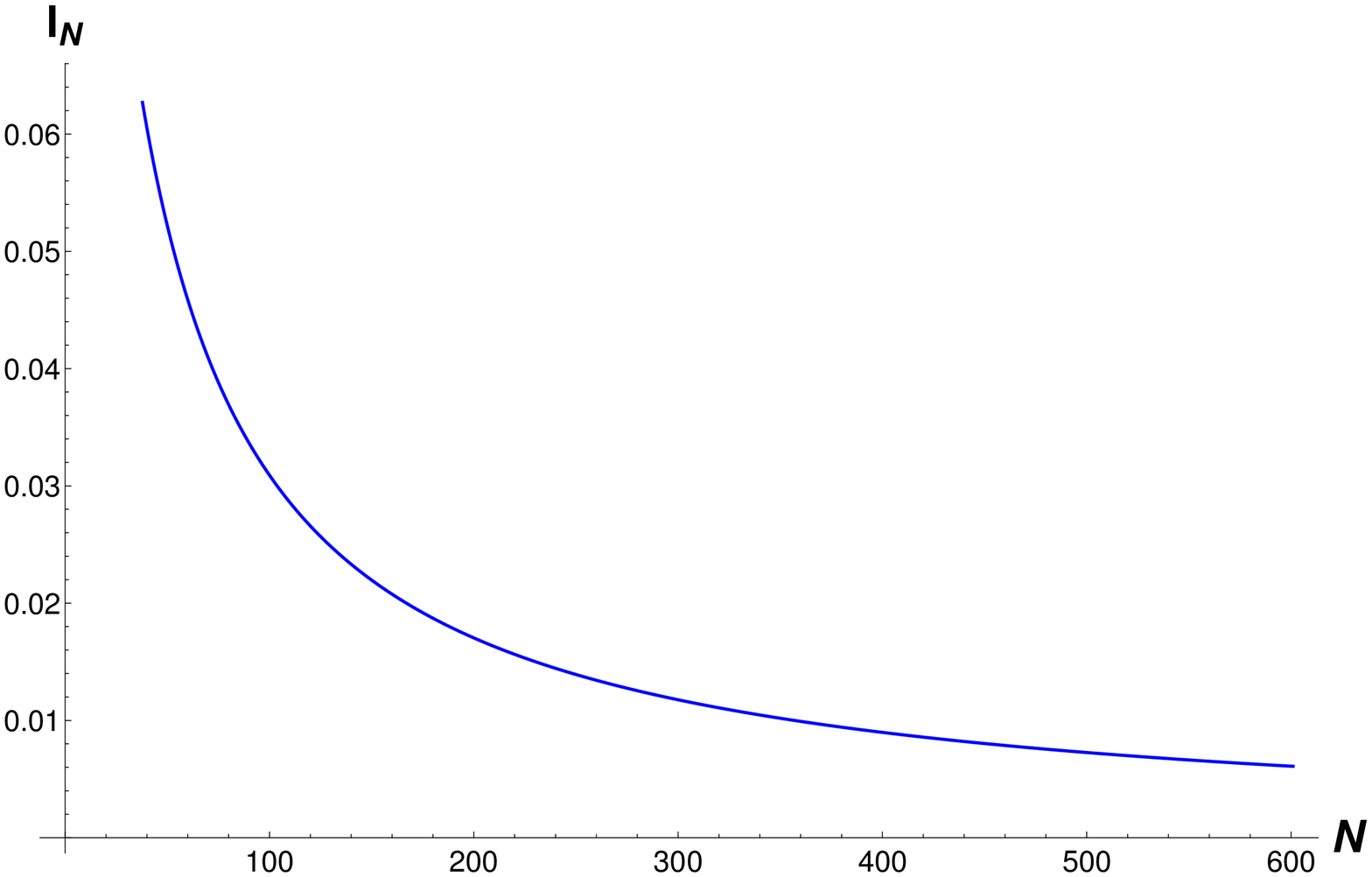}
 \hspace{3mm} 
\includegraphics[width=7cm]{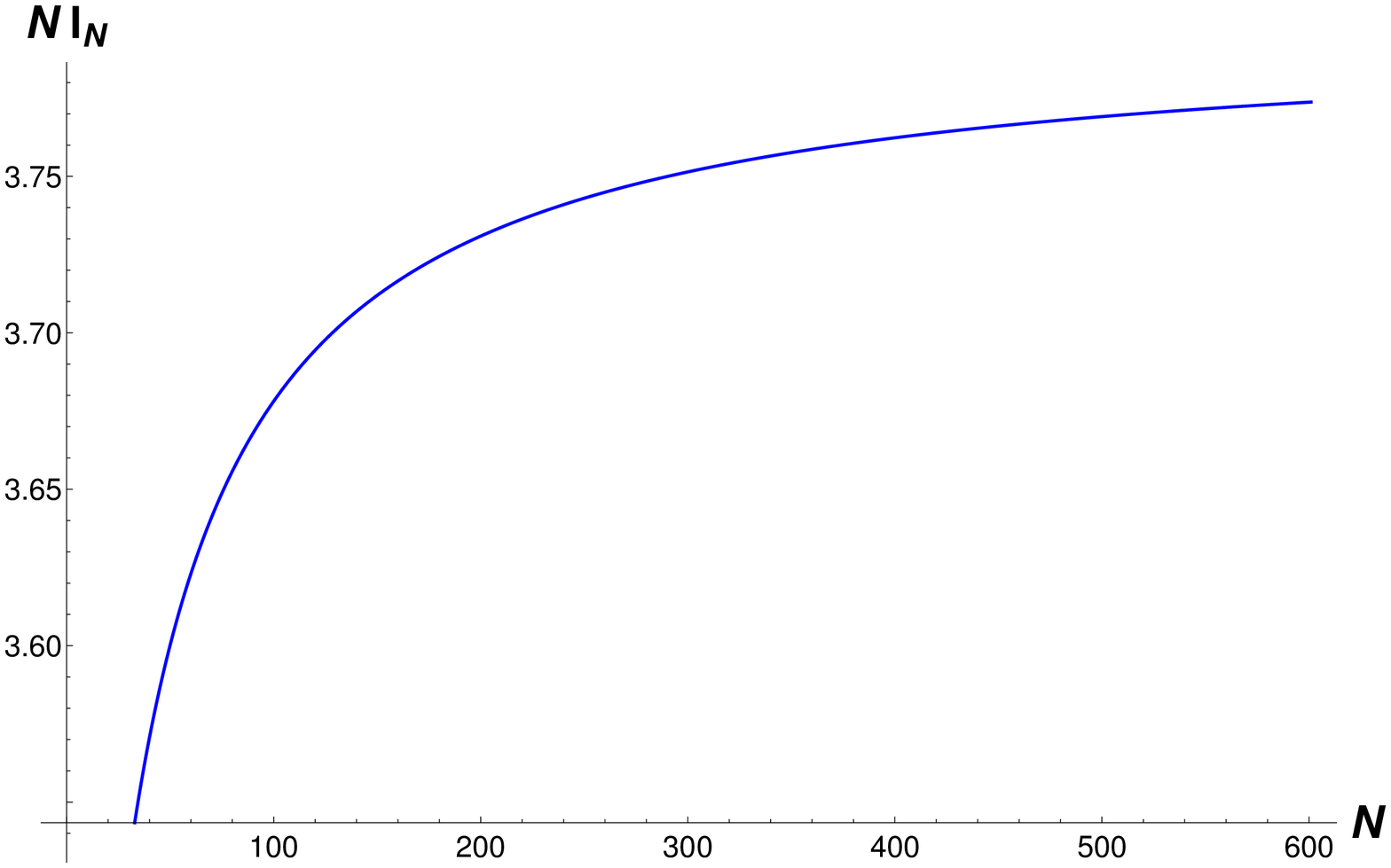}
\caption{
Integral $I^{}_N$ Eq.~(\ref{eq:In}) (left) and product $N I^{}_N$ (right) .
}
\label{fig:Fig5}
\end{figure}

To arrive at the equation of motion for the mean squared displacement, we assume that it is possible to determine the perturbative series up to a large order $N$.  
In this limit, using of the approximate formulas Eqs.~(\ref{eq:LadDen}) and (\ref{eq:LadNum}) is justified. 
If necessary, the limit $N\to\infty$ can be carried out at the end of calculations. We write Eq.~(\ref{eq:InMSqD}) as 
\begin{equation}
\label{eq:InMSqD2}
k^2  \langle r^2_{\mu}(0)\rangle =
\frac{\langle f^{(2)}_0\rangle+\langle\delta f^{(2)}\rangle}{\langle f^{(0)}_0\rangle+\langle\delta f^{(0)}\rangle}.
\end{equation}
In Eq.~(\ref{eq:InMSqD2}) $\langle f^{(i)}_0\rangle$ include all terms of the respective series with exception of the $N$th. 
The latter are contained in the fluctuation terms $\langle\delta f^{(i)}\rangle$ and are treated as the generators of the time 
evolution of $\langle r^2_{\mu}\rangle$ at small positive/negative times $\Delta t$. After some algebra presented in Appendix~\ref{app:ScalEq}
we obtain a self-consistent equation of motion for $\langle r^2_{\mu}(t)\rangle$ in the form of the second order ordinary differential equation 
\begin{equation}
\label{eq:DiffEq}
k^2\frac{\partial^2}{\partial t^2} \langle r^2_{\mu}(t)\rangle =  -E^2_g  I^{}_N\left[\frac{N}{2} - k^2 \langle r^2_{\mu}(t)\rangle\right]. 
\end{equation}
The integral $I^{}_N$ in the expression for the time-dependent diffusion coefficient is defined as
\begin{eqnarray}
\label{eq:In}
I^{}_N &=& \frac{1}{\langle f^{(0)}\rangle}\sum^{}_{s=\pm}\int^\infty_{-\infty} d\nu^{}_s ~ \nu^2_s~(2X^{}_s)^{2N} \approx 
\frac{\displaystyle \int^\infty_{-\infty} d\nu~ \nu^2 X^{2N}(\nu)}
{\displaystyle \int^\infty_{-\infty} d\nu~ X^{2N}(\nu)},
\end{eqnarray}
where we approximated $\langle f^{(0)}\rangle$ by the largest power in the series. Integral $I^{}_N$ shows a very slow decay for large $N$, cf. Fig.~\ref{fig:Fig5}, and seems 
to approach a finite limit for infinitely large $N$. Numerical routines struggle to compute it much beyond $N=600$ and it remains an open question whether the $N\to\infty$ bound exists. 
In turn, the product $N I^{}_N$ grows very slowly, which can also be interpreted as approaching a finite value in the limit $N\to\infty$.

The large-time asymptotics ($t\gg t^{}_c$, $t^{}_c$ being some crossover time far in the past, which can be chosen zero) is given by the solution of the self-consistent equation 
\begin{equation}
\label{eq:LargeT}
\langle r^2_{\mu}(t)\rangle^{>} \approx \frac{N}{2}\ell^2 + \ell^2{\cal C}\exp\left[-\frac{t}{\tau}\right],
\end{equation}
where we introduced the scattering time as 
\begin{equation}
\tau \approx \frac{1}{\sqrt{I^{}_N}E^{}_g} .
\end{equation}
In the limit $t\to\infty$ the mean squared displacement approaches its upper bound
\begin{equation}
\lim_{t\to\infty} \langle r^2_{\mu}(t)\rangle^{>} \to \frac{N}{2}\ell^2,
\end{equation}
which for $N\to\infty$ lies in the infinity and is therefore never reached. Hence it can be only approached from below, which requires $\cal C$ to be negative. 
If $\sqrt{I^{}_N}$ is small, then $\tau$ is large and the regime with linear time dependence should be broad. 
The diffusion coefficient is then obtained from 
\begin{equation}
\left.\frac{\partial}{\partial t} \langle r^2_{\mu}(t)\rangle^{>} \right|_{t=t^{}_c}= |{\cal C}|\frac{\ell^2}{\tau}. 
\end{equation}
Formally, $|{\cal C}|$ should follow from the initial condition at $t=t^{}_c$, but for this we need to know $\langle r^2_{\mu}(t^{}_c)\rangle^{>}$, 
which lies far in the past and therefore forgotten. In order to be a physical quantity, we demand for $D$ an invariance with respect to $N$. 
This is similar to the version of the renormalization group typically used in the high energy physics. This implies 
\begin{equation}
\frac{\partial}{\partial N} (\sqrt{I^{}_N}|{\cal C}|) = 0,
\end{equation}
from where then follows 
\begin{equation}
\sqrt{I^{}_N}|{\cal C}| = {\rm const}.
\end{equation}
Even though this constrain might appear not entirely transparent it has a natural analogy in the case of disordered electron gas without magnetic field.
Here, the diffusion coefficient is determined from the self-consistent Born approximation and appears unchanged in the partial series, e.g. cooperon or diffuson~[\onlinecite{Altshuler1994,Efetov1997}]. 
Comparison with Eq.~(\ref{eq:MsqD}) suggests this constant to be $2$. Then the physical diffusion coefficient becomes
\begin{equation}
\label{eq:DifCof}
D \approx \frac{E^{}_g}{k^2} \sim \ell l^{}_\lambda,
\end{equation}
i.e. it is proportional to the parametric volume of the model. 

Inserting the density of states from Eq.~(\ref{eq:DOS}) and the diffusion coefficient Eq.~(\ref{eq:DifCof}) into the Einstein relation Eq.~(\ref{eq:EinsteinRel}) yields the conductivity. 
The system is conducting within a parametric window located around each of the Landau zeros. 
The width of the conducting window is determined by the parameters of the microscopic model and by the disorder. 
The transition $g\to0$ is smooth and the conductivity degenerates to two sharp peaks at the Landau zeros. 
With increasing disorder, the peaks becomes broader and merge at some point to an amorphous structure. 
Simultaneously the amplitude becomes smaller, signaling the suppression of the conductivity in the strong disorder limit. 

We can compute the conductivity at an arbitrary Landau zero point. For weak disorder, the contribution from the other mode is negligible and we get for the density of states
\begin{equation}
\rho^{\rm LZ} \approx \frac{1}{\pi^{5/2}}\frac{k^2}{E^{}_g}.
\end{equation}
Using the units of $e^2/h$  instead of $e^2/\hbar$ adds an extra factor of $2\pi$, i.e.
\begin{equation}
\label{eq:CondLZ}
\sigma^{\rm LZ} \approx \frac{2\pi}{\pi^{5/2}} \frac{e^2}{h}  \approx 0.36 \frac{e^2}{h}.
\end{equation}
Numerically it is close to the dc limit of the optical conductivity of a single Dirac electron (which also has a zero energy point in the spectrum)
in clean and even weakly disordered systems, which depending on the way it is calculated is either $\frac{e^2}{\pi h} \approx 0.32\frac{e^2}{h}$ or 
$\frac{\pi e^2}{8h} \approx 0.39\frac{e^2}{h}$~[\onlinecite{Fradkin1986,Ludwig1994,Gusynin2006,Ziegler2006,Ziegler2007,Sinner2018}].
If compared to the established value evaluated for conventional electron gas with the spectrum without zero points, 
then the Padé-Borel resummation~[\onlinecite{Hikami1984a,Hikami1984b,Chakravarty1986}] gives the value $0.46\frac{e^2}{h}$, while the numerical
value of Ref.~[\onlinecite{Bhatt1993}] is $0.5\frac{e^2}{h}$ within a statistical uncertainty of 10\% .
However, we are not aware of any further results for the static conductivity at the Landau zero points.

\section{Discussions}

The electronic transport properties of disordered electronic systems in strong magnetic fields represents a technically very demanding problem, 
which evades a conclusive solution even despite several decades of active research. In the past there have been impressive break troughs, though. 
At the single-particle level, the problem has been effectively resolved by Ando~[\onlinecite{Ando1974}] and Wegner~[\onlinecite{Wegner1983}].
The latter work provided an exact expression for the single-particle propagator. The resulting Green's function does not reveal any singularities and 
describes a state of the matter without pronounced resonances and consequently without clearly defined quasiparticles. 

On the technical side the difficulties aggregate if one goes beyond the single-particle picture and considers processes involving two or more particles. 
While the single-particle propagator retains its form in the position space in every order of perturbative expansion, which suggests a kind of invariance 
under the operation of perturbative expansion, the two-particles propagator does not show up this property. 
Instead, the disorder washes out the expansion series in the position space, making higher order terms broader in comparison to the clean system.
This is the reason, why the Wegner's technique cannot be applied to the two-particles propagator with the same success. 
From studying previous works available to us, it is not clear, whether this simple fact has been realized so far. 

In our work we perform the perturbative expansion in powers of disorder strength and classify all scattering processes into two main groups: 
Firstly those which dress the single-particle propagators. To order $g^3$ all of these diagrams are shown in Fig.~\ref{fig:Fig7}.  
They are absorbed into the fully dressed single-particle propagators, for which exact results are known~[\onlinecite{Wegner1983,Brezin1984}]. 
Secondly, there are processes which cannot be accounted for at the single-particle level. They are represented by loop diagrams, 
which involve at least one scattering process between the advanced and retarded sides. To order $g^3$ all of them are shown in Fig.~\ref{fig:Fig6}.
We take all diagrams to third order to into account, without truncating the perturbative series. 
Ideally, one would compute all diagrams in all orders of expansion. 
The number of diagrams grows rapidly with increasing order of expansion though, while the computational effort for the evaluation of every individual diagram increases considerably. 
However, it is possible to extract the diagrammatic channel which dominates the large-scale behavior of the two-particles propagator.
Still, due to absence of a small expansion parameter, the obtained series do not show any sign of convergence~[\onlinecite{Aoki1987}]. 

Our main intention is to calculate the transport coefficients, in particular the static conductivity. Because the series diverge, a naive use of the Kubo formula fails.  
Therefore we approach the conductivity via the Einstein relation, which requires the knowledge of the density of states and of the diffusion coefficient. 
While the former is known from the Wegner's solution, the latter is not. We acquire it from the mean-squared displacement, which at large times should be proportional to the diffusion coefficient. 
For infinite times the mean-squared displacement approaches an infinite time independent value. 
The processes linear in time emerge at the time scales determined by the combined action of disorder and magnetic field. 
The former being weak and the latter strong puts this characteristic time to considerable absolute values, which can indeed appear as the large time limit in an experiment.  
Whatever happens at larger times can evoke an impression of the onset subdiffusion.
The infinite times asymptotics might be inaccessible in realistic measurements. 
With this we can compute the conductivity from the Einstein relation with the extracted diffusion coefficient.

For discussions of a concrete physical system we consider the BZH-Hamiltonian~[\onlinecite{Zhang2006}]. 
Its lowest Landau level energy spectrum has two modes, corresponding to the two particle species related to each other by the time-reversal symmetry. 
Both modes exhibit zero energy points at critical magnetic fields, the so-called Landau zero modes.
It is possible to lift the degeneracy of the Landau zero points parametrically, i.e. to move them apart from each other on the magnetic field axis.
The analytical expressions for the conductivity adopted for this case  suggest that the system becomes metallic in the parametric window around the Landau zeros. 
The weaker the disorder, the narrower is the metallic window. This picture appears rather intuitive and hopefully likely to be confirmed experimentally with a modest effort.
Adjusting the magnetic field exactly to the zero Landau energy gives for the conductivity a universal value $\sim 0.36e^2/h$, which is surprisingly close to the established 
results for the conductivity of the disordered Dirac electrons without magnetic fields, which also have zero energy points in the spectrum.

The most obvious avenue leading beyond the scope of the present work is the challenge of reconciling the Einstein relation with the Kubo-Greenwood formalism~[\onlinecite{Wegner1979b,McKane1981,Ludwig1994,Sinner2018}]. 
In the Kubo-Greenwood formalism, the computation of the conductivity requires the knowledge of the two-particles propagators as well.
However, usually the Kubo-Greenwood formula does not involve any normalization and the issue of series convergence becomes much more urgent. 
Curiously though, the static conductivity of the clean system can be straightforwardly evaluated from the Kubo-Greenwood formula, giving the value of $e^2/2h$ for each Landau mode.
At the first superficial glance, once the disorder is brought into the system this conductivity is destroyed. 
Another possible direction is an extension to the higher Landau levels. 
These problems are left for the future activities. 

\section{Acknowledgments} 
A.S. acknowledges the support by the grants of the Julian Schwinger Foundation for Physics Research 
and expresses his gratitude to Prof. C. Weber for hospitality during the later phase of manuscript preparation. 
The work was concluded at IMDEA Nanotechnologia Madrid with the support from the personal research grant (A.S.)
PCI2021-122057-2B (MSCA  IF  EF-ST  2020) of the Agencia Estatal de Investigacion, Spain.

\appendix

\section{BHZ-Hamiltonian in explicit form}
\label{app:Ham}
 
Without magnetic field, the BHZ-Hamiltonian Eq.~(\ref{eq:BHZH}) reads:
\begin{eqnarray}
& H = & \\
\nn
&\displaystyle
\left(
\begin{array}{cccc}
\displaystyle -{T^{}_-}{\nabla^{}_+\nabla^{}_-} - \epsilon^{}_0 + \Delta^{}_0  &  - iv\nabla^{}_-  &  0  &  0  \\
  - iv\nabla^{}_+                                        &\displaystyle -{T^{}_+}{\nabla^{}_+\nabla^{}_-} - \epsilon^{}_0 - \Delta^{}_0 & 0 & 0 \\
   0 & 0 & \displaystyle -{T^{}_-}{\nabla^{}_+\nabla^{}_-} - \epsilon^{}_0 + \Delta^{}_0 &   - iv\nabla^{}_+   \\
   0 & 0 &  - iv\nabla^{}_-  & \displaystyle -{T^{}_+}{\nabla^{}_+\nabla^{}_-} - \epsilon^{}_0 -  \Delta^{}_0
\end{array}
\right).
&
\end{eqnarray}
All quantities are explained in the Section~\ref{Sec:Ham}. The model does not presume any coupling or scattering mechanisms between spin projections (or valleys), 
which would appear in the off-diagonal blocs. To introduce the magnetic field we replace 
$\displaystyle \nabla^{}_{-} \to 2\partial^{}_z + k^2\bar z    = A$, $\nabla^{}_{+} \to  2\partial^{}_{\bar z} - k^2 z = A^\dag$, and 
$\displaystyle \nabla^{}_{+}\nabla^{}_{-} \to (2\partial^{}_{\bar z} - k^2 z)( 2\partial^{}_z + k^2\bar z ) - 2k = A^\dag A - 2k$.

\begin{figure}[t]
\includegraphics[height=6.6cm]{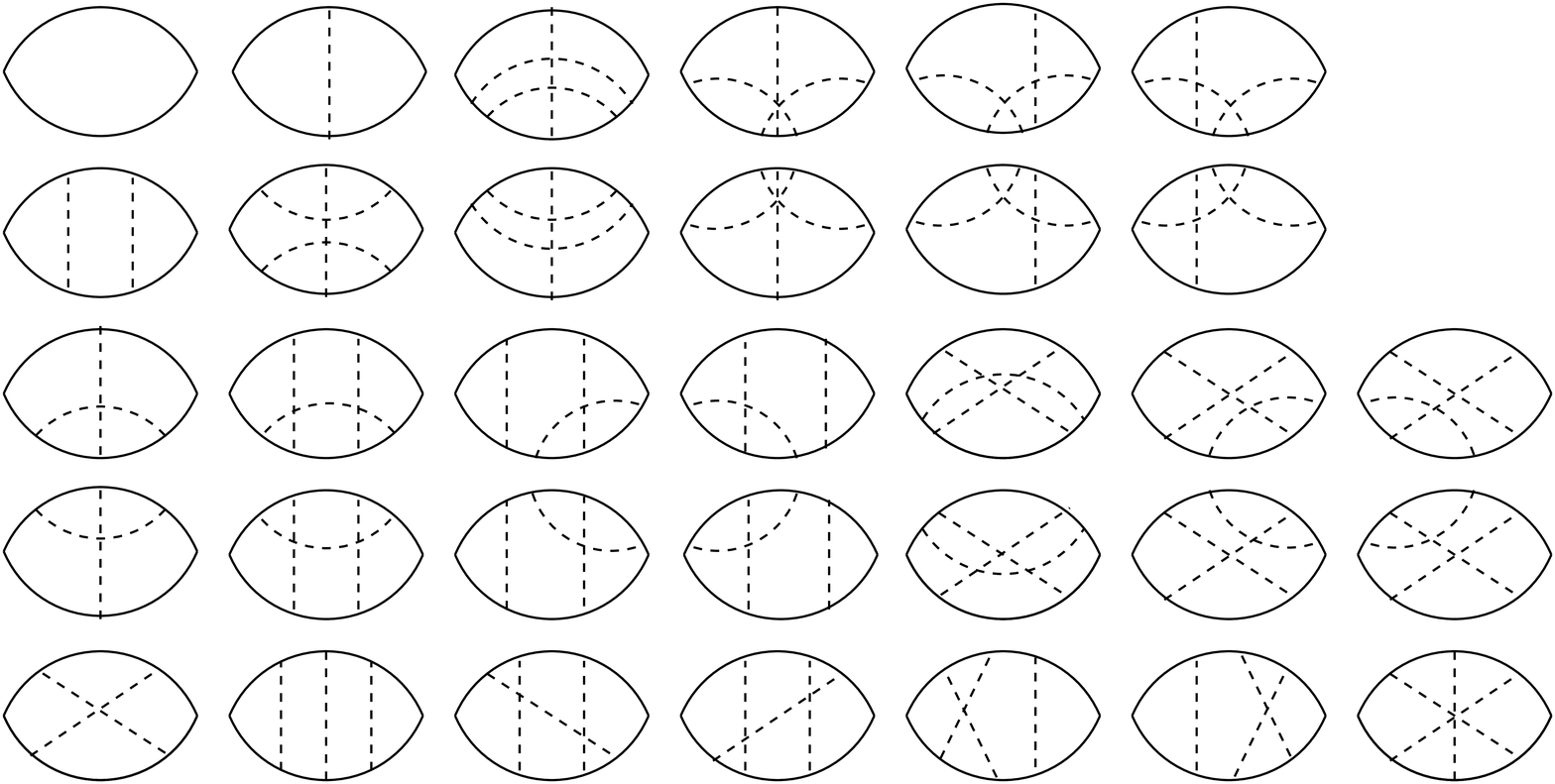}
\caption{Perturbative processes contributing to the dressing of the two-particles propagator up to the third order in disorder strength. 
Solid lines denote the fully dressed Wegner's propagators and the dashed lines the disorder correlators.} 
\label{fig:Fig6}
\end{figure}

\section{Technical preliminary} 
\label{app:teciss}

In Section~\ref{Sec:OptCon} we emphasize that, while the position dependence of the single-particle propagator $G^{\pm}_{rr'}$ of an electron in lowest Landau level is preserved 
in every order of perturbative expansion, this is not so for the two-particles propagator $G^{+}_{rr'}G^{-}_{r'r}$. 
We demonstrate this by evaluating respective leading order terms. The calculations are easy to perform using the following integration formula
\begin{equation}
\label{eq:MasterEq}
\frac{1}{\pi} \int d\bar x dx ~ e^{\displaystyle -[c \bar xx - a\bar yx-b\bar xz ]} = \frac{1}{c} \exp\left[\frac{ab}{c}\bar yz\right], \;\;\; (a,b,c) > 0 .
\end{equation}

After performing the disorder average in the lowest order of the single-particle propagator we have
\begin{equation}
\langle G^{\pm}_{r x} v^{}_{x}G^{\pm}_{x y}v^{}_{y}G^{\pm}_{y r'} \rangle^{}_g 
= g\sum^{}_{xy}\delta^{}_{xy}G^{\pm}_{rx} G^{\pm}_{xy} G^{\pm}_{yr'}
= g\sum^{}_{x}G^{\pm}_{rx} G^{\pm}_{xx} G^{\pm}_{xr'}.
\end{equation}
The local propagator $G^{\pm}_{xx}$ is a position independent constant. 
Because of the idempotence of projectors, the product of Green's functions is separable in the spin space  
(for sake of simplicity we skip below the convergence factor $\pm i0^+$, unless the opposite is necessary):
\begin{eqnarray}
g\sum^{}_{x}G^{\pm}_{rx} G^{\pm}_{xx} G^{\pm}_{xr'} = g\left(\frac{k^2}{2\pi}\right)^3 
\sum_{s=\pm}
\frac{{\cal P}^{}_s}{[E - E^{}_s]^3}
\int d\bar xd x~
e^{-\frac{k^2}{2}(|r|^2+|x|^2 - 2\bar rx)}
e^{-\frac{k^2}{2}(|x|^2+|r'|^2 - 2\bar xr')}.
\end{eqnarray}
We evaluate the integral as follows:
\begin{eqnarray}
\nn
\int d\bar xd x~
e^{-\frac{k^2}{2}(|r|^2+|x|^2 - 2\bar rx)}
e^{-\frac{k^2}{2}(|x|^2+|r'|^2 - 2\bar xr')} &=& e^{-\frac{k^2}{2}(|r|^2+|r'|^2)}\int d\bar xd x~e^{-k^2( \bar xx - \bar rx - \bar xr')} \\
&=& e^{-\frac{k^2}{2}(|r|^2+|r'|^2-2\bar r r')}
,
\end{eqnarray}
where in the last line we used  the formula Eq.~(\ref{eq:MasterEq}). 
Putting all terms together we get 
\begin{equation}
 g\sum_{x}G^{}_{rx} G^{}_{x x} G^{}_{x r'} 
= \frac{g}{2}\left(\frac{k^2}{2\pi}\right)^2 e^{-\frac{k^2}{2}(|r|^2+|r'|^2-2\bar r r')}
\sum_{s=\pm} \frac{{\cal P}^{}_s}{[E - E^{}_s]^3}  .
\end{equation}
Indeed, the spatial part of this expression is the same as that of the Green's function of the clean system. 
This property can be traced back in every order of perturbative expansion, irrespective of the diagram's topology. 
This means that the disorder does not affect the coherence of the single-particle Green's function and therefore, 
the series can be summed up exactly~[\onlinecite{Wegner1983,Brezin1984}].

To illustrate the effect of the disorder on the two-particles propagator we start with the expression of the clean system:
\begin{equation}
\label{eq:2PGFcl}
 G^{+}_{rr'}G^{-}_{r'r} = \left(\frac{k^2}{2\pi}\right)^2 e^{-k^2(\bar r-\bar r')(r-r')} \sum_{s=\pm} \frac{{\cal P}^{}_s}{(E-E^{}_s-i0^+)(E-E^{}_s+i0^+)}.
\end{equation}
The two-particles propagator is a real valued quantity and decays exponentially in the position space with the Gaussian exponent 1.
The leading order perturbative vertex correction reads 
\begin{eqnarray}
g\sum_x  G^{+}_{rx}G^{+}_{xr'}G^{-}_{r'x}G^{-}_{xr} &=& g \left(\frac{k^2}{2\pi}\right)^4\sum_{s=\pm} \frac{{\cal P}^{}_s}{(E-E^{}_s-i0^+)^2(E-E^{}_s+i0^+)^2} \\
&\times&\int d\bar xdx ~ e^{-k^2(\bar r-\bar x)(r-x)}e^{-k^2(\bar x-\bar r')(x-r')}
\end{eqnarray}
At first, the integral can be rewritten as 
\begin{equation}
\int d\bar xdx ~ e^{-k^2(\bar r-\bar x)(r-x)}e^{-k^2(\bar x-\bar r')(x-r')} = e^{-k^2(|r|^2 + |r'|^2)}
\int d\bar xdx ~ e^{-k^2[2|x|^2 -\bar x(r+r') - (\bar r + \bar r')x]} ,
\end{equation}
which gives with the help of Eq.~(\ref{eq:MasterEq})
\begin{equation}
g\sum_x  G^{+}_{rx}G^{+}_{xr'}G^{-}_{r'x}G^{-}_{xr} = 
\frac{g}{2^2}\left(\frac{k^2}{2\pi}\right)^3 e^{-\frac{k^2}{2}(\bar r-\bar r')(r-r')}  \sum_{s=\pm}  \frac{{\cal P}^{}_s}{(E-E^{}_s-i0^+)^2(E-E^{}_s+i0^+)^2} .
\end{equation}
Comparing this with Eq.~(\ref{eq:2PGFcl}) we recognize the different Gaussian exponent 1/2, 
i.e. the spatial part of the first correction is broader than that of the clean system. 
This effect takes place in all orders but differently for topologically different diagrammatic channels,
i.e. the coherence of the two-particles propagator is violated by the disorder.
This is the ultimate reason why the adoption of the Wegner's technique to the two-particles propagator is difficult. 
Table~\ref{tab:Tab1} shows the Gaussian exponents for each diagram from Fig.~\ref{fig:Fig6}. 
They reveal certain patterns for particular diagrammatic channels. 
For instance, one recognizes that the large scale behavior in every order of perturbative expansion is dominated by the diagrams belonging 
to the so-called ladder channel. Those are the diagrams no. 1,2,3 and 4. 
The dominant oscillating channel is represented by the diagrams no. 5, 7, etc., which we call the menora diagrammatic channel. 
The diagrams belonging to the so-called fan channel, e.g. no. 4, 17, etc., retain the same Gaussian exponent in every perturbative order. 
In Appendix~\ref{app:MSD} we argue that they contribute the least to the amplitudes in every order of expansion and therefore are clearly subdominant in comparison to the ladder channel.

\begin{table}
\begin{tabular}{ccccccccccc}
No. &  Diagram  &  Exponent & \hspace{2mm} &  No.  & {\rm Diagram } &  Exponent & \hspace{2mm} & No. & {\rm Diagram} & Exponent  \\
\\
1
&
\includegraphics[height=3mm]{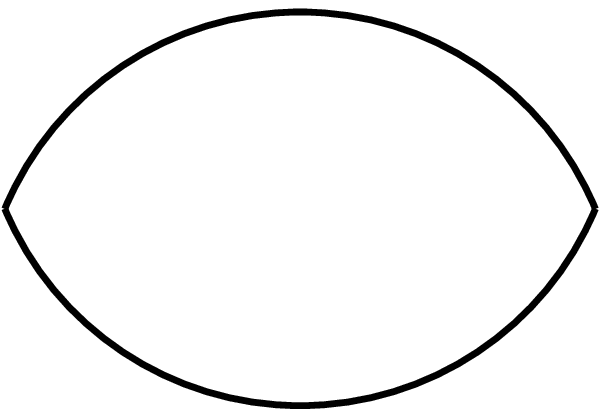} 
&  
$1$
&
&
6
& 
\includegraphics[height=3mm]{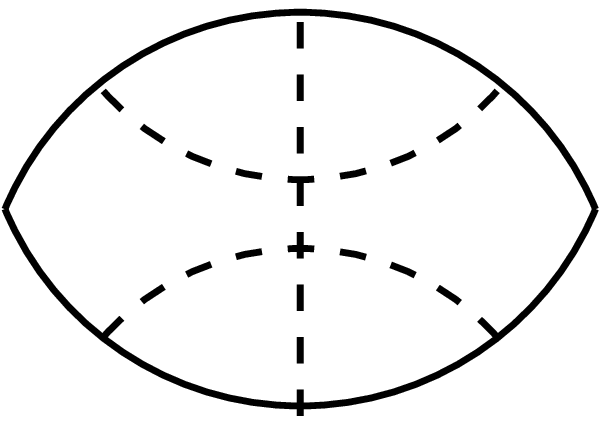} 
&
$\frac{3}{4}$
&
&
12
&
\includegraphics[height=3mm]{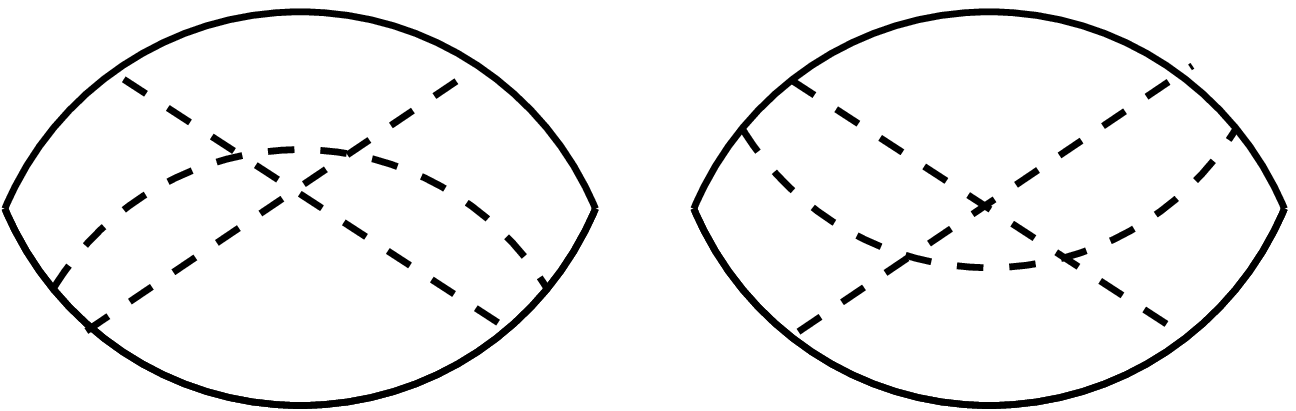}
&
$\frac{2}{3}$
\\
\\
2
&
\includegraphics[height=3mm]{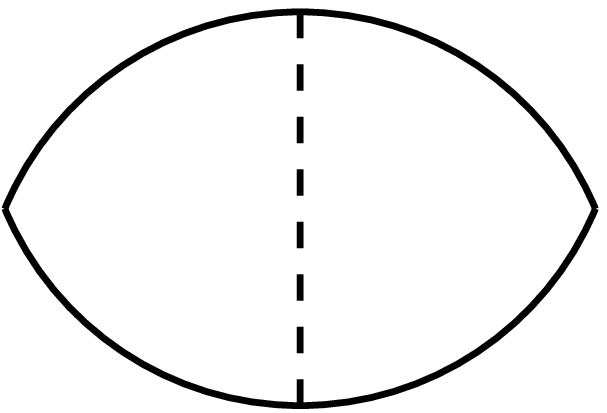}
& 
$\frac{1}{2}$
&
&
7
&
\includegraphics[height=3mm]{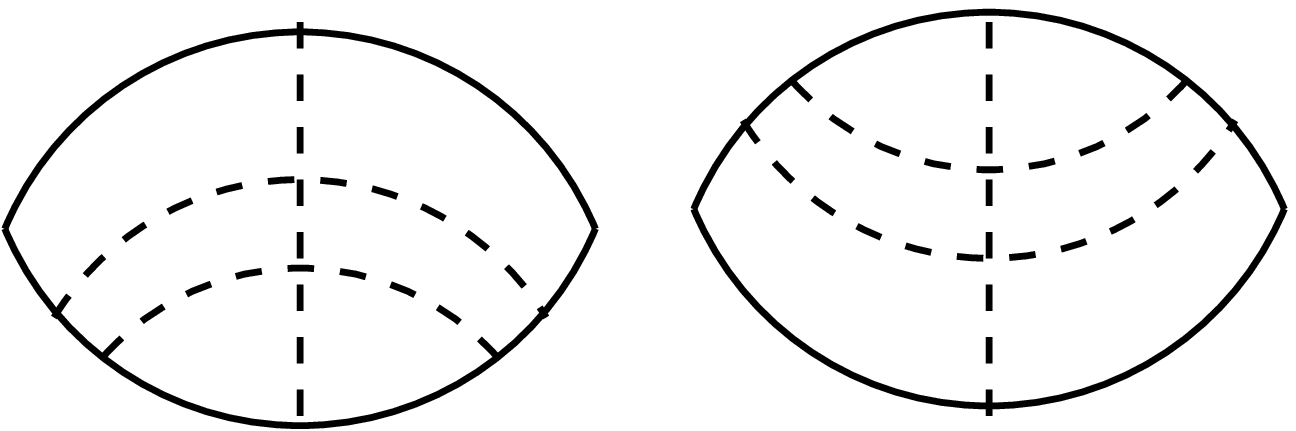} 
&
$\frac{3}{4}$
&
&
13
&
\includegraphics[height=7mm]{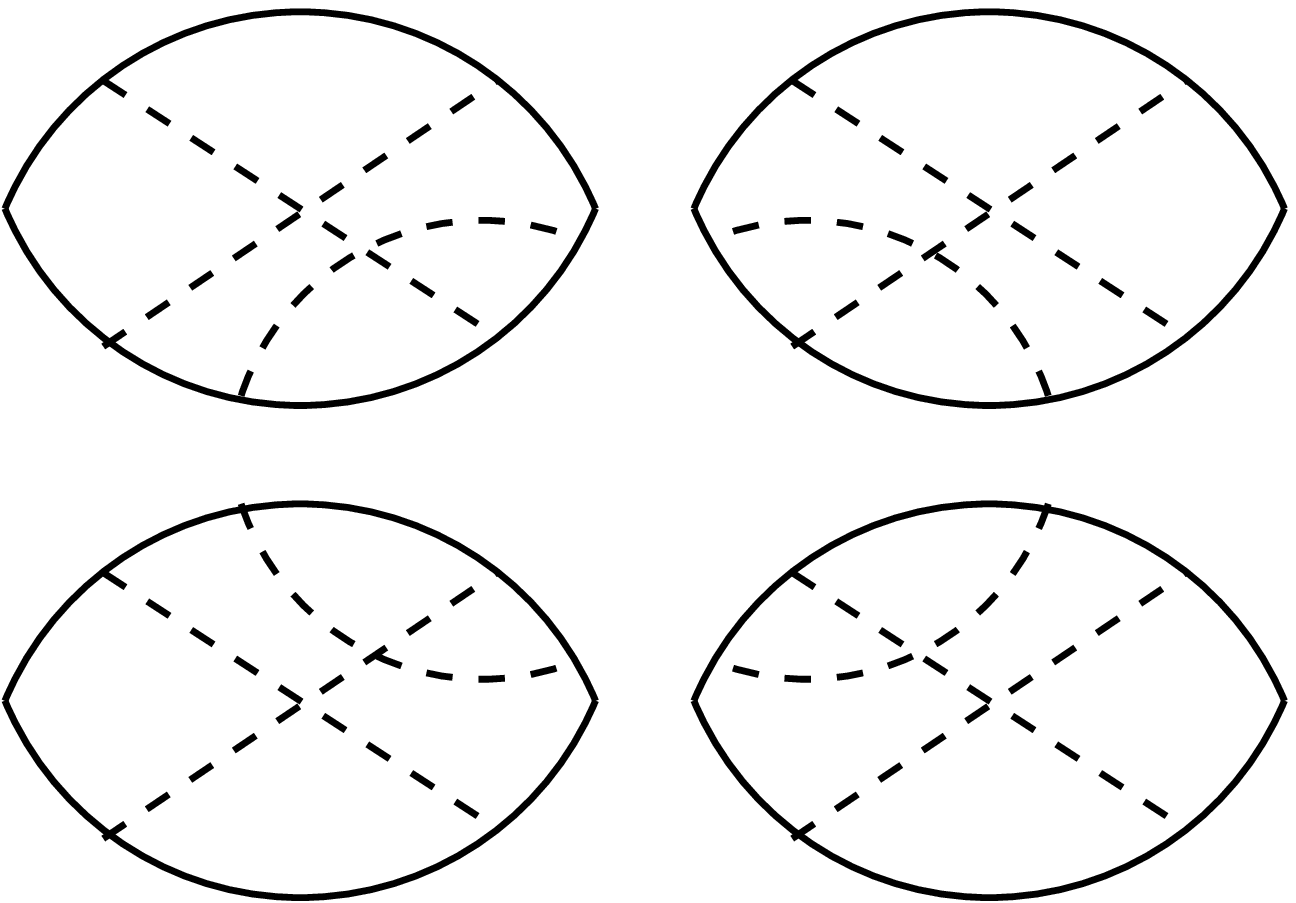} 
&
$\frac{1}{2}$
\\
\\
3
&
\includegraphics[height=3mm]{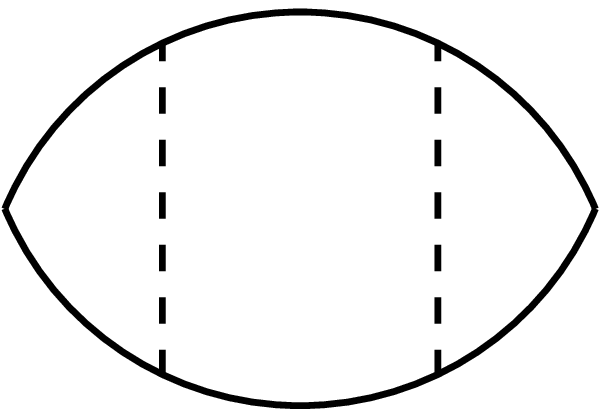} 
&
$\frac{1}{3}$
&
&
8
&
\includegraphics[height=3mm]{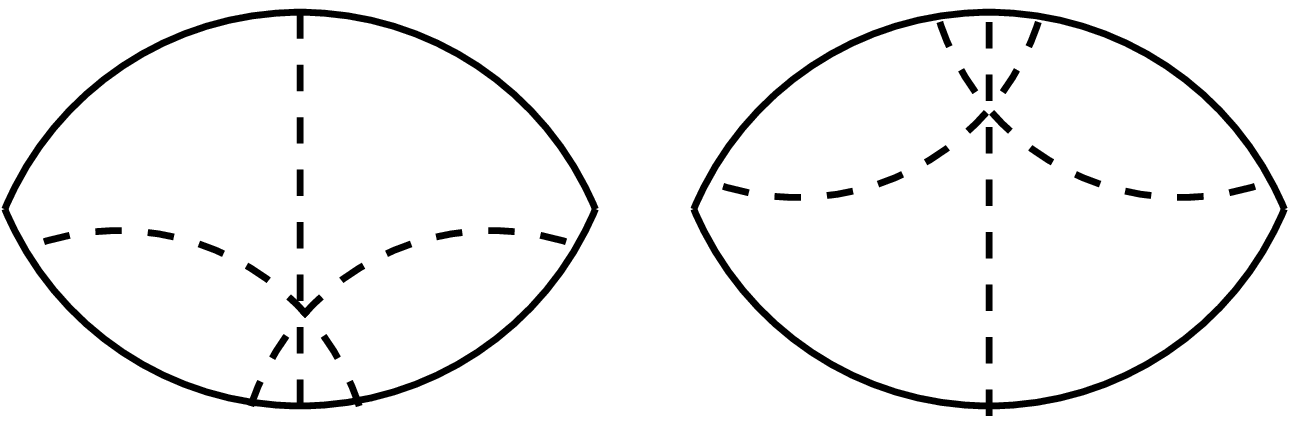}
&
$\frac{2}{3}$
&
&
14
&
\includegraphics[height=3mm]{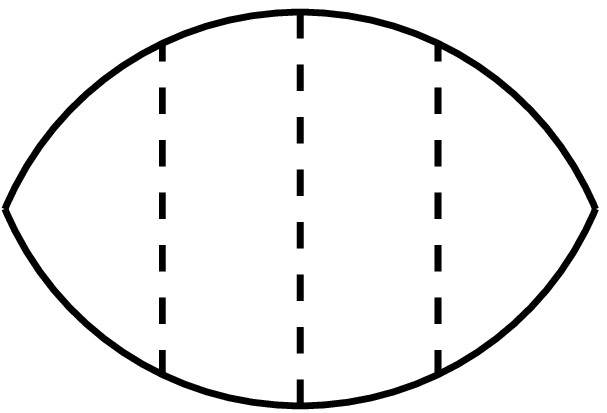} 
&
$\frac{1}{4}$
\\
\\
4
&
\includegraphics[height=3mm]{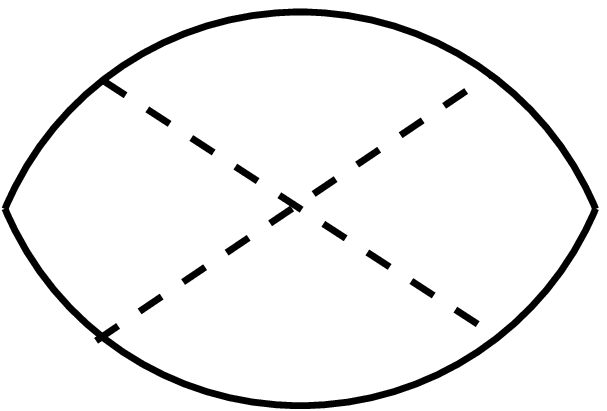} 
&
$\frac{1}{2}$
&
&
9
&
\includegraphics[height=7mm]{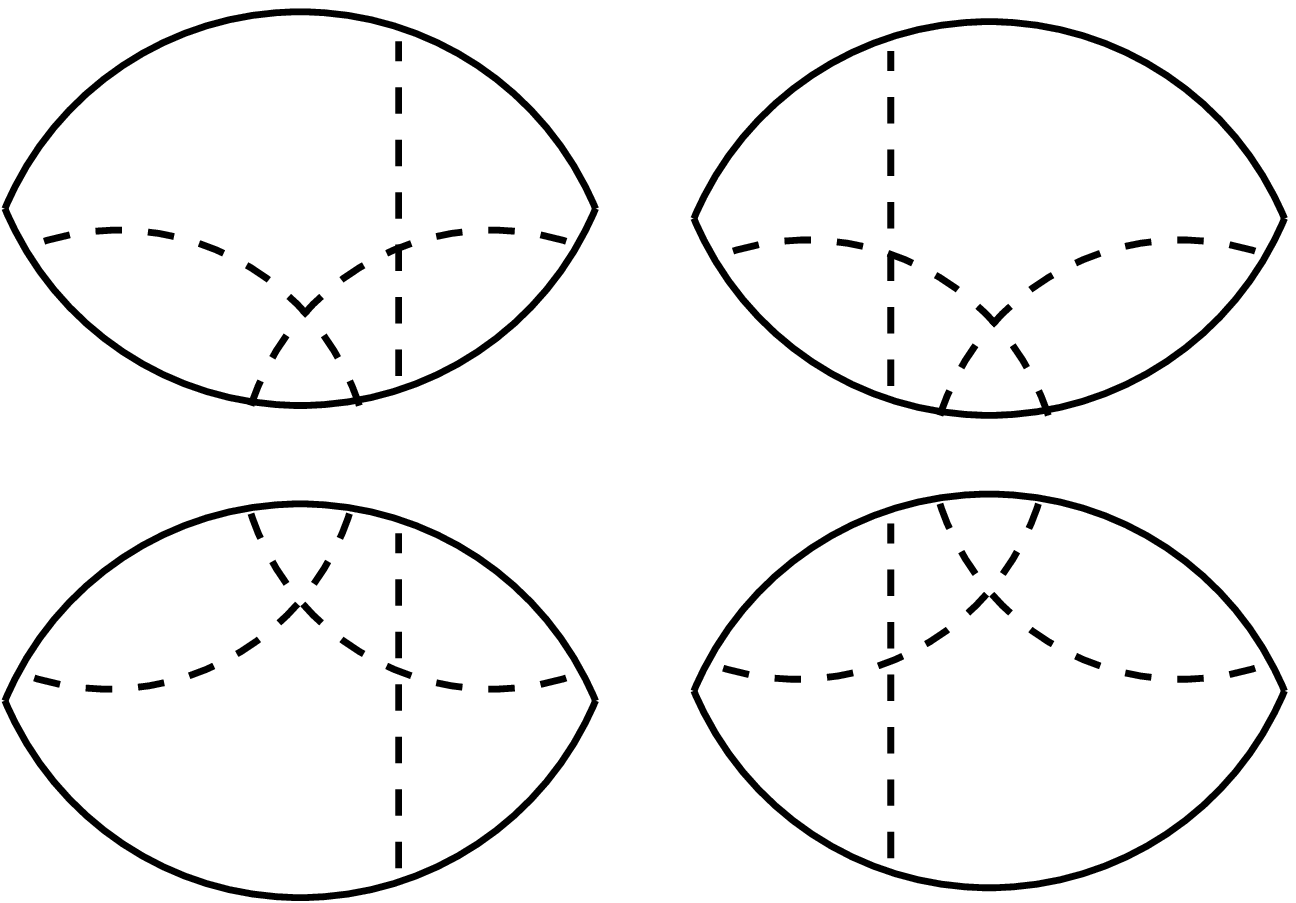}
&
$\frac{3}{5}$
&
&
15
&
\includegraphics[height=3mm]{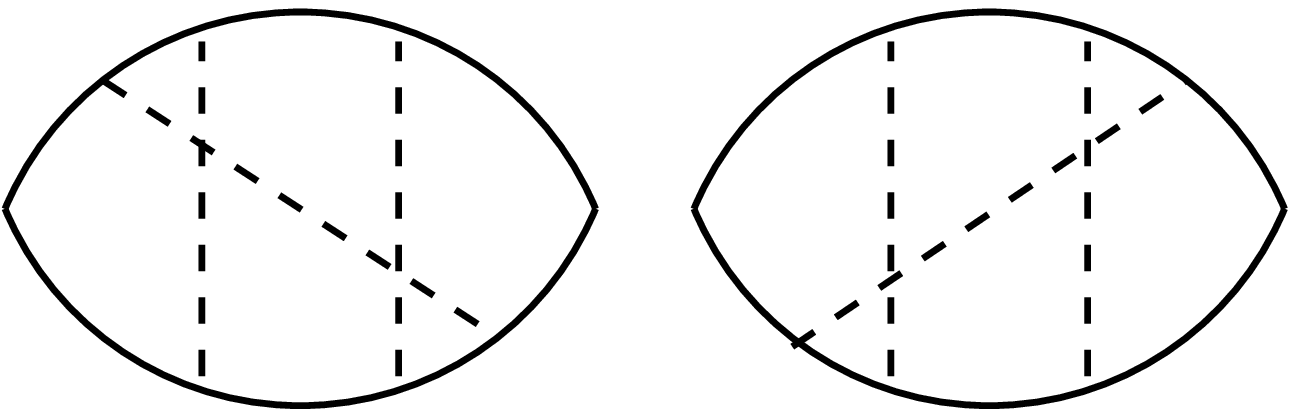} 
&
$\frac{1}{2}$
\\
\\
5
&
\includegraphics[height=3mm]{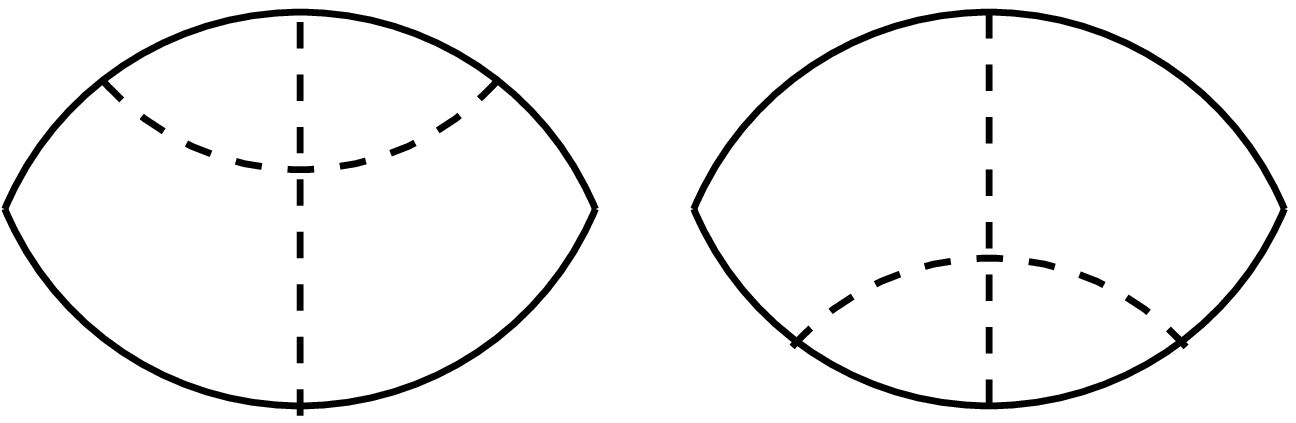} 
&
$\frac{2}{3}$
&
&
10
&
\includegraphics[height=3mm]{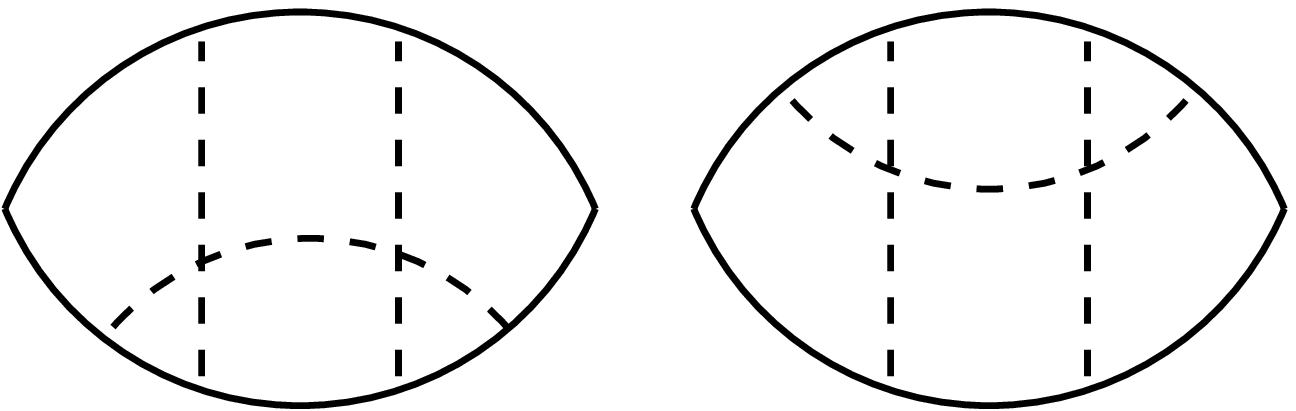}
&
$\frac{3}{5}$
&
&
16
&
\includegraphics[height=3mm]{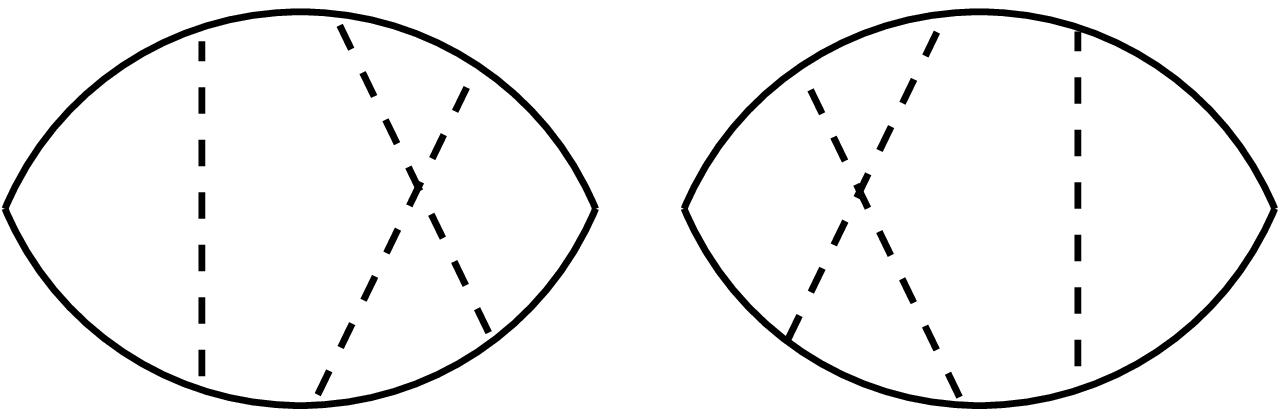}
&
$\frac{1}{3}$
\\
\\
&
&
&
&
11
&
\includegraphics[height=7mm]{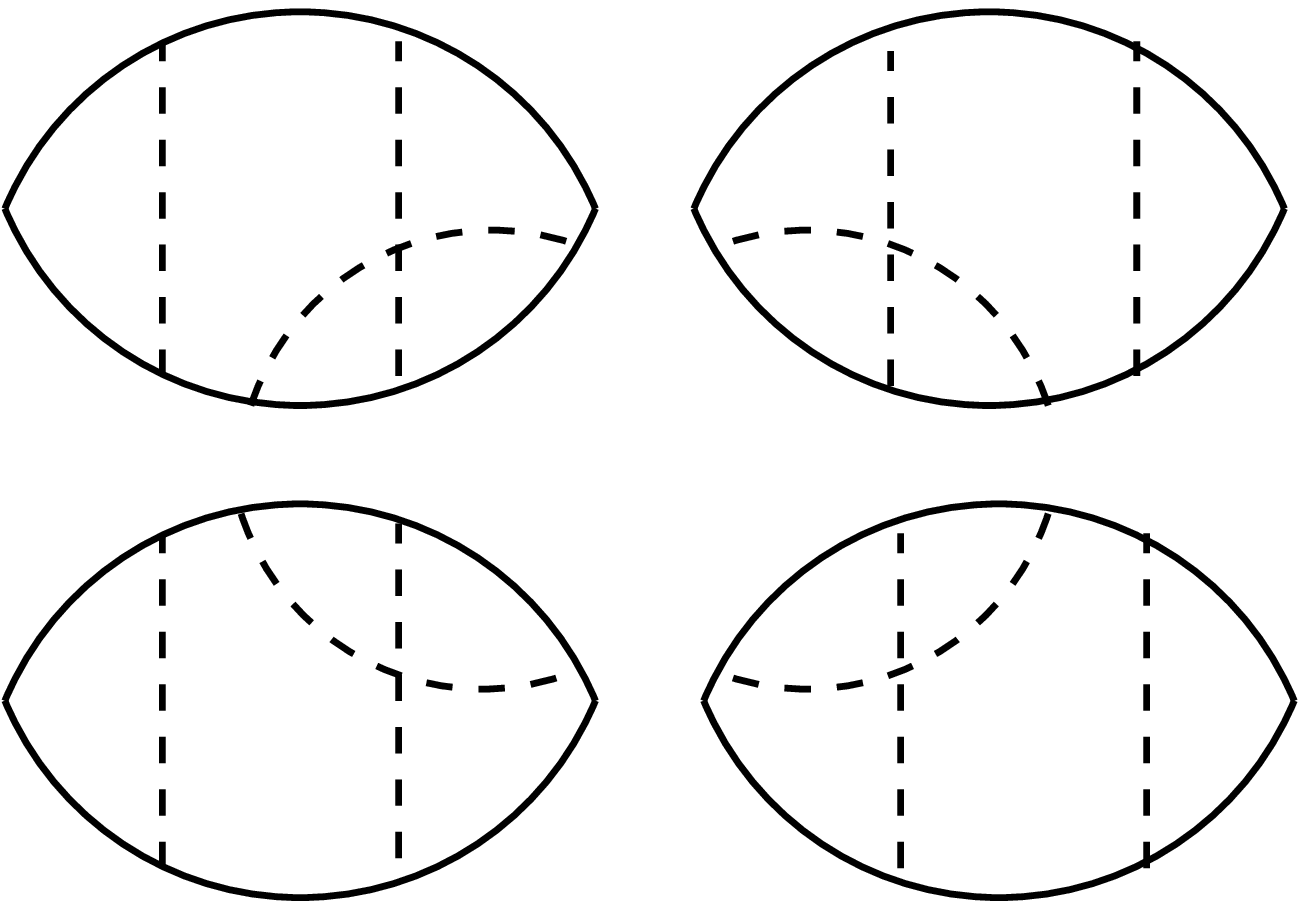}
&
$\frac{2}{5}$
&
&
17
&
\includegraphics[height=3mm]{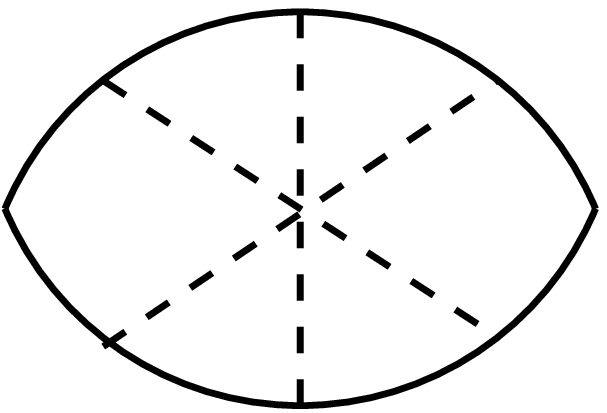}
&
$\frac{1}{2}$
\end{tabular}
\caption{
Gaussian exponents $\chi^{}_i$ of all perturbative processes contributing to the two-particles Green's function shown in Fig.~\ref{fig:Fig6}. 
}
\label{tab:Tab1} 
\end{table}

\section{Exact Wegner's propagator of the disordered system}
\label{app:WegnerGF}

\begin{figure}[t]
\includegraphics[height=6cm]{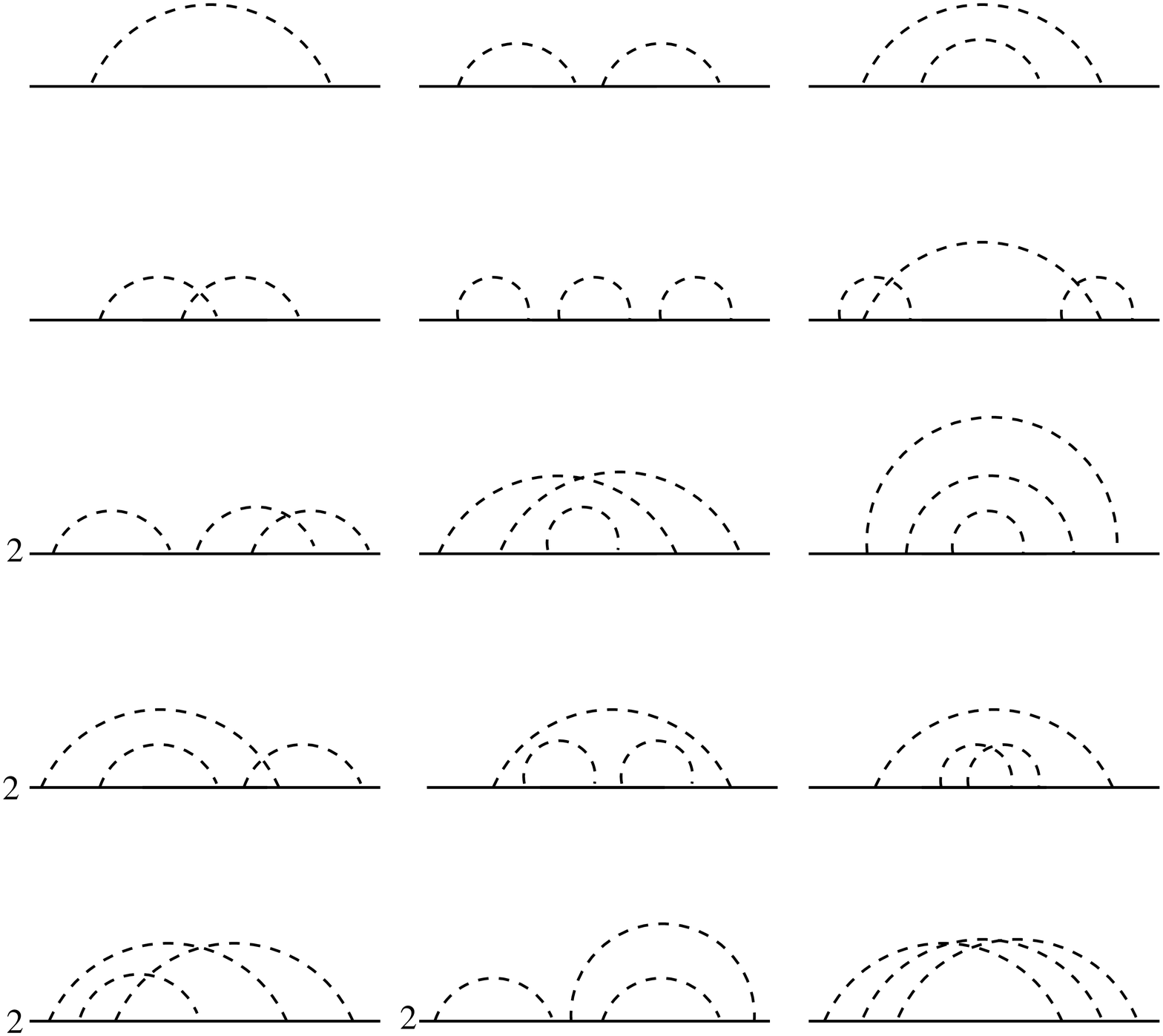}
\caption{
Perturbative processes contributing to the dressing of the single-particle propagator due to the disorder to order $g^1$ (one diagram), $g^2$ (three diagrams), and $g^3$ (fifteen diagrams).
}
\label{fig:Fig7}
\end{figure}

We start the discussion of the dressing effect for the single-particle propagator by evaluating all perturbative contributions to order $g^2$ and $g^3$.
The number of diagrams to order $g^n$, $n$ being a positive integer number and zero is 
$\frac{(2n)!}{2^n n!}$, hence, there are in total 3 diagrams to the order $g^2$ and 15 diagrams to order $g^3$, Fig.~\ref{fig:Fig7}. 
In Table~\ref{tab:Tab2} we summarize their amplitudes. 
Next order $g^4$ would require the evaluation of 105 diagrams with a larger computational 
effort for each individual diagram. To this end though, we have an asymptotically exact expression:
\begin{equation}
\label{eq:ExAs}
\bar G^{\pm}_{rr'} (E) =  \frac{k^2}{\pi}e^{-\frac{k^2}{2}(|r|^2+|r'|^2-2\bar r r')}\sum^{}_{s=\pm}{\cal P}^{}_s {\cal F}^{\pm}_s(E), 
\end{equation}
where the frequency dependent part of the Green's function reads 
\begin{eqnarray}
\label{eq:exp1}
{\cal F}^{\pm}_s(E) &=& \frac{1}{2}
\frac{1}{E - E^{}_{s}}
\left[
1 + \frac{E^2_g}{[E - E^{}_{s}]^2} 
+ \frac{5}{2}\frac{E^4_g}{[E - E^{}_{s}]^4}  
+ \frac{37}{4}\frac{E^6_g}{[E - E^{}_{s}]^6}\cdots
\right], 
\end{eqnarray}
where $E^2_g=\frac{gk^2}{4\pi}$. Not surprisingly, the expansion coefficients 1, 1, 5/2, 37/4.. are precisely those of the exact solution by 
Wegner Ref.~[\onlinecite{Wegner1983}], extended by Br\'{e}zin et al. with a different technique in Ref.~[\onlinecite{Brezin1984}].

\begin{table}
\begin{tabular}{ccccccccccc}
No. &  Diagram  &  Amplitude & \hspace{2mm} &  No.  & {\rm Diagram } &  Amplitude & \hspace{2mm} & No. & {\rm Diagram} & Amplitude   \\
\\
1
&
\includegraphics[height=4mm]{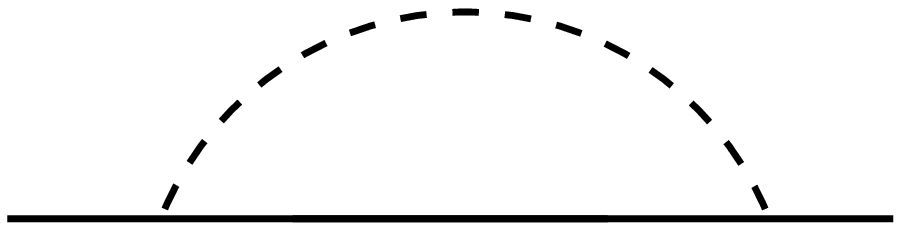} 
&  
${\displaystyle\sum_s} E^{}_g F^3_s$
&
&
6
& 
\includegraphics[height=4mm]{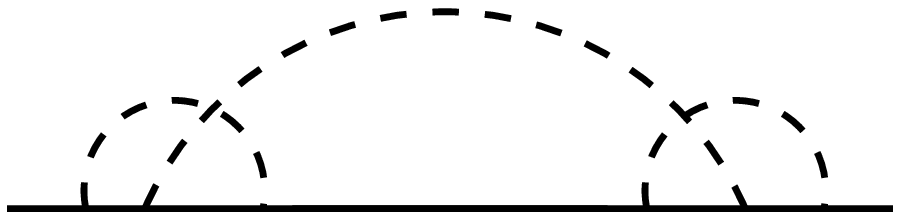} 
&
$\frac{1}{3}{\displaystyle\sum_s}E^3_g F^7_s$
&
&
11
&
\includegraphics[height=4mm]{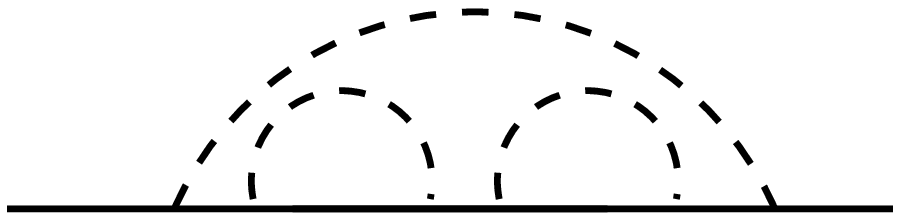}
&
${\displaystyle\sum_s}E^3_g F^7_s$
\\
\\
2
&
\includegraphics[height=3mm]{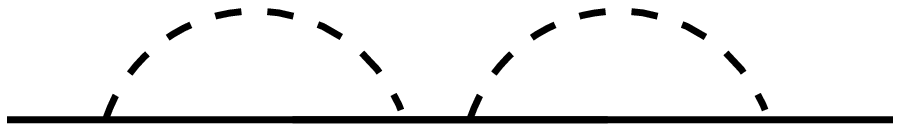}
& 
${\displaystyle\sum_s}E^2_g F^5_s$
&
&
7
&
\includegraphics[height=3.5mm]{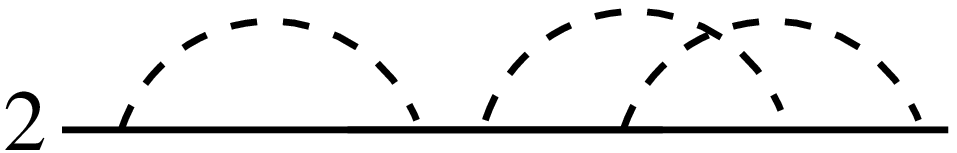} 
&
${\displaystyle\sum_s} E^3_g F^7_s$
&
&
12
&
\includegraphics[height=4.mm]{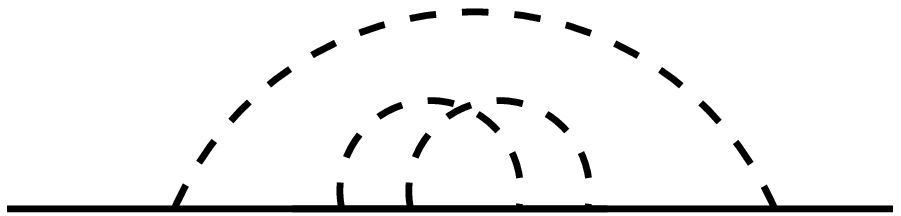} 
&
$\frac{1}{2}{\displaystyle\sum_s} E^3_g F^7_s$
\\
\\
3
&
\includegraphics[height=4mm]{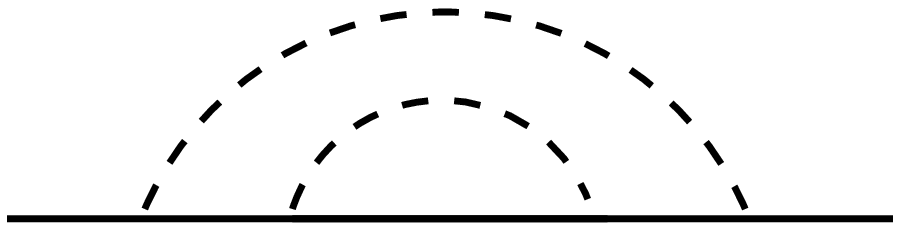} 
&
${\displaystyle\sum_s} E^2_g F^5_s$
&
&
8
&
\includegraphics[height=4mm]{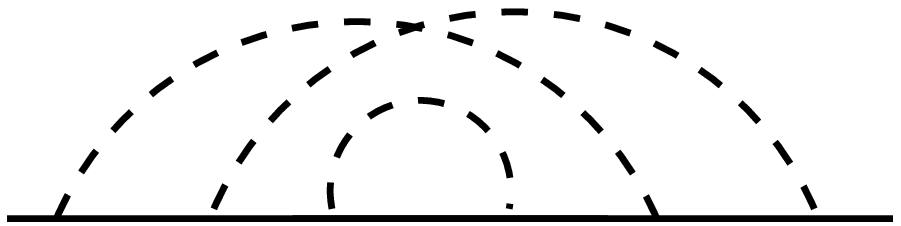}
&
$\frac{1}{2}{\displaystyle\sum_s}E^3_g F^7_s$
&
&
13
&
\includegraphics[height=5mm]{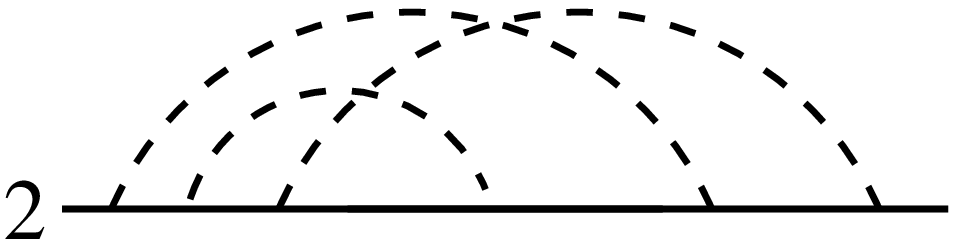} 
&
$\frac{2}{3}{\displaystyle\sum_s} E^3_g F^7_s$
\\
\\
4
&
\includegraphics[height=3mm]{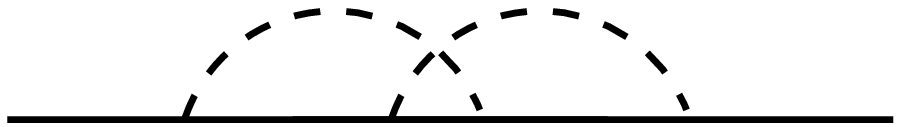} 
&
$\frac{1}{2}{\displaystyle\sum_s} E^2_g F^5_s$
&
&
9
&
\includegraphics[height=7mm]{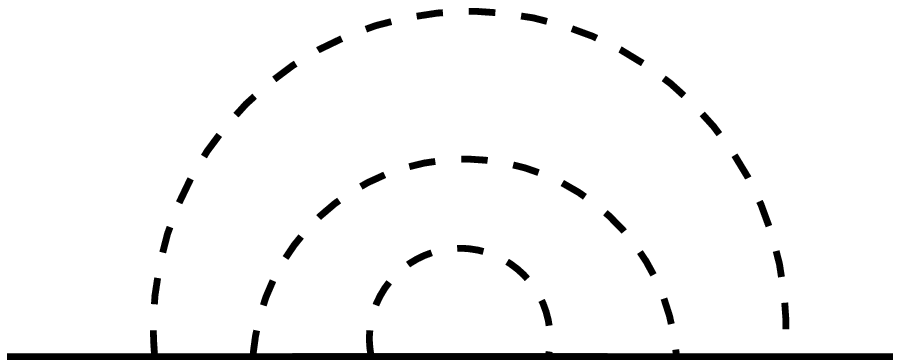}
&
${\displaystyle\sum_s} E^3_g F^7_s $
&
&
14
&
\includegraphics[height=5mm]{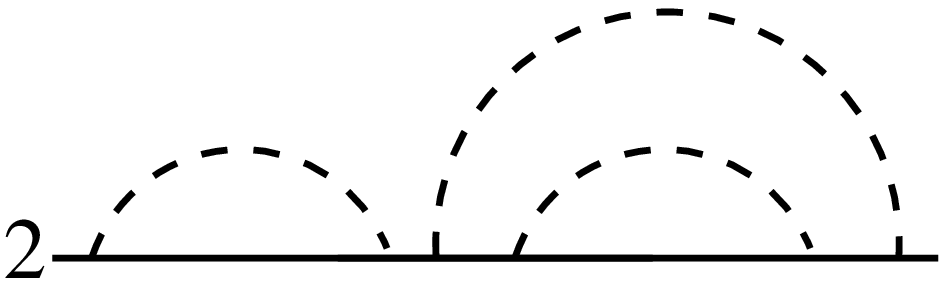} 
&
$2{\displaystyle\sum_s}E^3_g F^7_s $
\\
\\
5
&
\includegraphics[height=3mm]{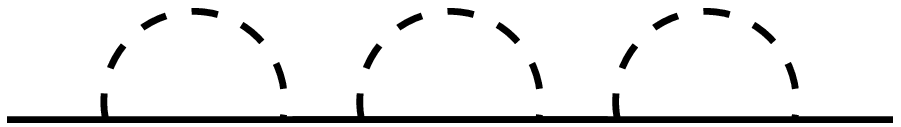} 
&
${\displaystyle\sum_s} E^3_g F^7_s $
&
&
10
&
\includegraphics[height=5mm]{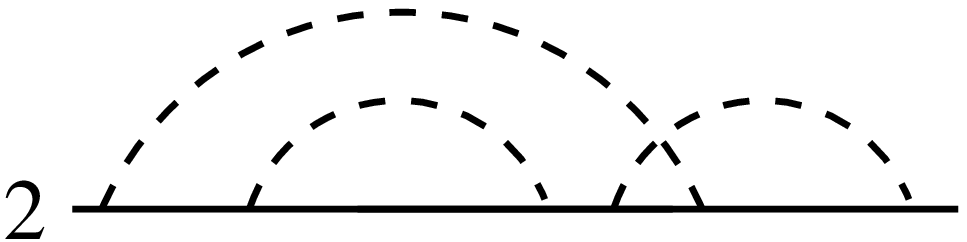}
&
${\displaystyle\sum_s}E^3_g F^7_s $
&
&
15
&
\includegraphics[height=4mm]{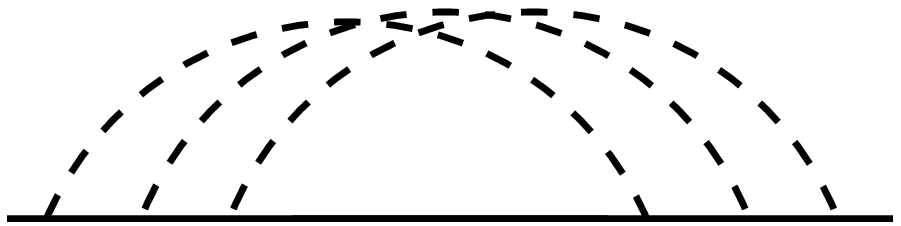}
&
$\frac{1}{4}{\displaystyle\sum_s}E^3_g F^7_s $
\end{tabular}
\caption{Amplitudes of each perturbative process contributing to the dressing of the single-particle propagator
up to the order $g^3$ in multiples of $k^2/2\pi$.}
\label{tab:Tab2} 
\end{table}

\begin{figure}[t]
\includegraphics[width=7cm]{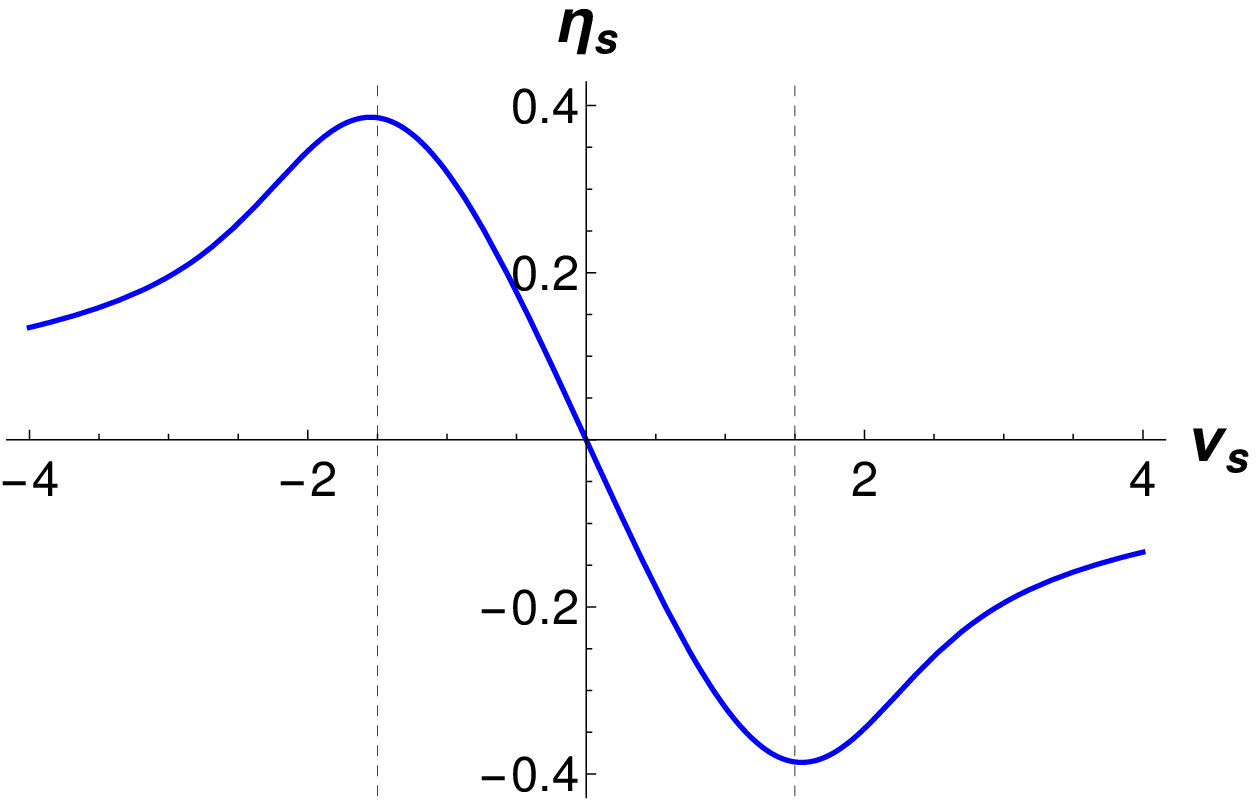}
\hspace{5mm}
\includegraphics[width=7cm]{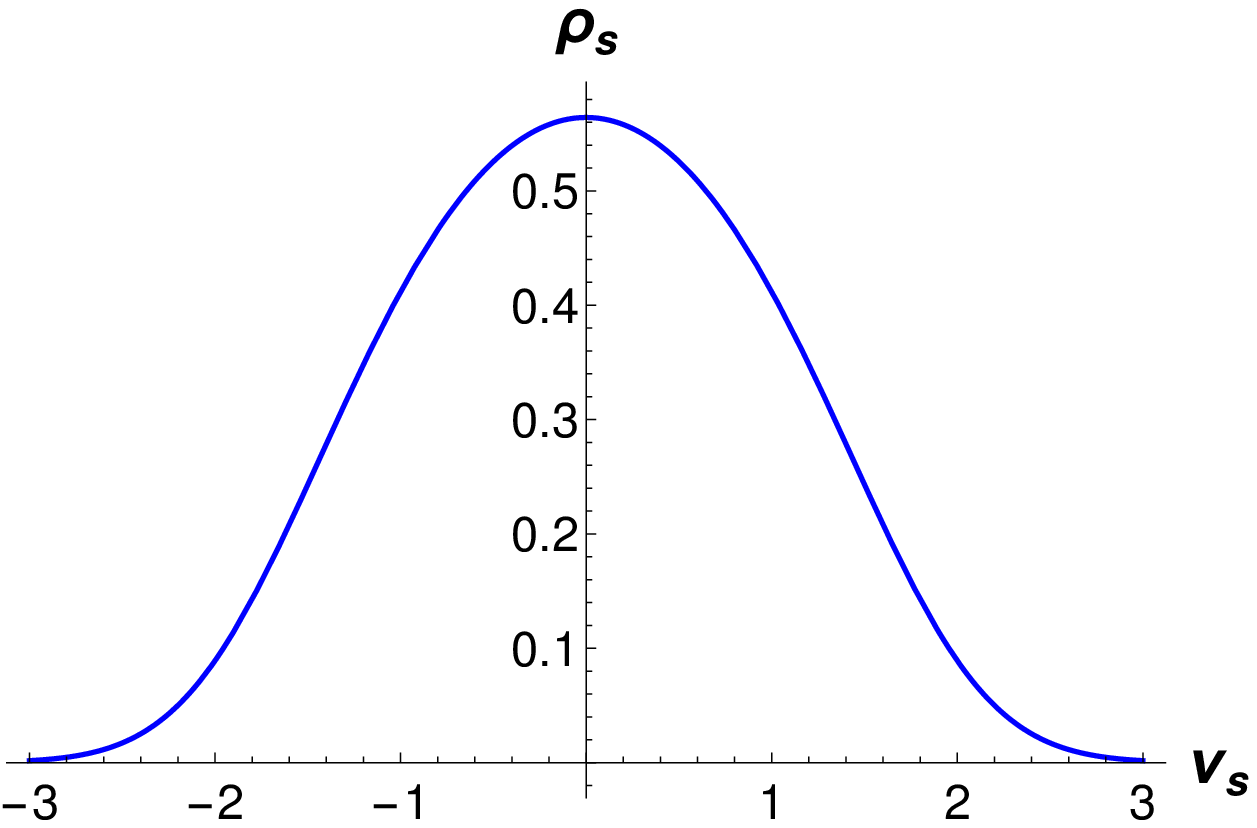}
\caption{
The shape of the real (left) and imaginary (right) part of the Wegner's propagator. 
}
\label{fig:Fig8}
\end{figure}

Following the line of Wegner's reasoning, the series in the rectangular brackets in Eq.~(\ref{eq:exp1}) is related to the zero dimensional functional integral
\begin{eqnarray}
\label{eq:Wegner}
\frac{1}{a}\sum^\infty_{n=0} f^{}_n\left(-\frac{b}{a^2}\right)^n 
&=& 
\frac{\displaystyle \int d\phi^\ast d\phi~ \phi^\ast\phi ~ e^{-a\phi^\ast\phi-\frac{b}{4}(\phi^\ast\phi)^2}}
{\displaystyle \int d\phi^\ast d\phi~ e^{-a\phi^\ast\phi-\frac{b}{4}(\phi^\ast\phi)^2}}, 
\end{eqnarray}
with $f^{}_n$ representing the combinatorial factor. In order to insure the convergence, the real part of $a$ has to be positive.
The integral can be carried out exactly by quadratic complement, giving
\begin{eqnarray}
\frac{1}{a}\sum^\infty_{n=0} f^{}_n\left(-\frac{b}{a^2}\right)^n 
&=& 
-\frac{\partial}{\partial a}
\log\left[
\frac{2\pi}{\sqrt{b}} e^{-\frac{a^2}{b}}\int^\infty_{\frac{a}{\sqrt{b}}} dt~e^{-t^2}
\right].
\end{eqnarray}
The integral under the logarithm is rewritten as
\begin{equation}
 \int^\infty_{\frac{a}{\sqrt{b}}} dt~e^{-t^2} = \frac{\sqrt{\pi}}{2}\left[1 -\frac{2}{\sqrt{\pi}} \int^{\frac{a}{\sqrt{b}}}_0 dt~e^{-t^2}\right],
\end{equation}
and therefore 
\begin{eqnarray}
\frac{1}{a}\sum^\infty_{n=0} f^{}_n\left(-\frac{b}{a^2}\right)^n 
&=& 
-\frac{\partial}{\partial a}
\log\left[
e^{-\frac{a^2}{b}}\left(1 -\frac{2}{\sqrt{\pi}} \int^{\frac{a}{\sqrt{b}}}_0 dt~e^{-t^2}\right)
\right] \\
&=& \frac{2}{\sqrt{b}}
\left[
\frac{a}{\sqrt{b}} + \frac{1}{\sqrt{\pi}}
\frac{\displaystyle e^{-\frac{a^2}{b}}}{\displaystyle 1 -\frac{2}{\sqrt{\pi}} \int^{\frac{a}{\sqrt{b}}}_0 dt~e^{-t^2}}
\right].
\end{eqnarray}
Plugging into this expression  
\begin{equation}
\label{eq:ab}
 a = \mp i \left[E  - E^{}_{s} \pm i0^+\right],\;\;\; {\rm and} \;\;\; b = E^2_g,
\end{equation}
we get the frequency dependent part of the propagator for each spectral mode of the lowest Landau level
\begin{equation}
 {\cal F}^{\pm}_s(E) = -\frac{1}{2E^{}_g}
 \left[
 2\nu^{}_s \pm \frac{i}{\sqrt{\pi}} \frac{e^{\nu^2_s}}{\displaystyle 1 \mp i \frac{2}{\sqrt{\pi}}\int^{\nu^2_s}_0 dt ~ e^{t^2}}
 \right],
\end{equation}
where $\nu^{}_s = (E-E^{}_s)/E^{}_g$. From here one obtains the real and imaginary part of the Green's function Eqs.~(\ref{eq:WRE}) and (\ref{eq:WIM}) shown in Fig.~\ref{fig:Fig8}. 
The imaginary part, which is essentially the contribution to the net density of states from each spectral mode in the lowest Landau level, 
is placed symmetrically around the Landau zero point with the maximum at the band center. 

In order to Fourier transform the Wegner's propagator with respect to the energy we start with Eq.~(\ref{eq:exp1}) rewritten in accord with Eq.~(\ref{eq:Wegner}) 
\begin{eqnarray}
\label{eq:eq:exp2}
{\cal F}^{\pm}_s(E) & = & \mp\frac{i}{2}
\frac{1}{\mp i(E - E^{}_{s})}
\left[
1 - \frac{E^2_g}{[\mp i(E - E^{}_{s})]^2} 
+ \frac{5}{2}\frac{E^4_g}{[\mp i(E - E^{}_{s})]^4}  
- \frac{37}{4}\frac{E^6_g}{[\mp i(E - E^{}_{s})]^6}\cdots
\right].
\end{eqnarray}
Then, the leading order term is easily Fourier transformed with the help of the Cauchy theorem 
\begin{equation}
\mp\int\frac{dE}{2\pi i} \frac{e^{\pm iE t}}{E - E^{}_{s} \pm i0^+} = e^{\pm iE^{}_s t},
\end{equation}
with the contour integration stretching over the entire complex plane, hence necessary enveloping the pole.
All higher terms follow essentially by taking derivatives of the oscillating function on the right hand side with respect to $E^{}_s$, e.g.
\begin{eqnarray}
\int\frac{dE}{2\pi} \frac{e^{\pm iE t}}{[\mp i(E - E^{}_{s} \pm i0^+)]^{2n+1}} &=& \mp \frac{(-1)^n}{2n!}\frac{\partial^{2n}}{\partial E^{2n}_s} \int\frac{dE}{2\pi i}\frac{e^{\pm iE t}}{E - E^{}_{s} \pm i0^+} \\
&=& 
\frac{(-1)^n}{2n!}\frac{\partial^{2n}}{\partial E^{2n}_s}e^{\pm iE^{}_s t} = \frac{t^{2n}}{2n!}e^{\pm iE^{}_s t}.
\end{eqnarray}
This gives for the Fourier transformed Wegner's function the following series
\begin{equation}
\label{eq:TWegnerEsr}
{\cal F}^{\pm}_s(t) = \mp\frac{i}{2}e^{\pm iE^{}_s t}\left[1 - \frac{(E^{}_gt)^2}{2!} + \frac{5}{2}\frac{(E^{}_gt)^4}{4!} - \frac{37}{4} \frac{(E^{}_gt)^6}{6!} \cdots \right].
\end{equation}
The combinatorial factors of the series in the rectangular brackets can be determined to every order using the integral representation Eq.~(\ref{eq:Wegner}). We therefore can write it as 
\begin{equation}
\label{eq:WGFFT}
 {\cal F}^{\pm}_s(t) = \mp\frac{i}{2}e^{\pm iE^{}_s t}\Omega(E^{}_gt), \;\;\; \Omega(E^{}_gt) = \sum^\infty_{n=0} (-1)^nf^{}_n\frac{(E^{}_gt)^{2n}}{(2n)!},
\end{equation}
where $f^{}_n$ denotes the combinatorial factors from Eq.~(\ref{eq:Wegner}).

\section{Evaluation of the mean squared displacement: Technical issues}
\label{app:MSD}

After accounting for all processes which dress the single-particle Green's propagators, the disorder effect for the two-particles propagator and 
correspondingly for the mean squared displacement comes from the loop diagrams, which involve at least one scattering process between advanced and retarded sides. 
All such diagrams to order $g^3$ are shown in Fig.~\ref{fig:Fig6}. 
Every single diagram can be evaluated exactly in a very efficient way using the master formula Eq.~(\ref{eq:MasterEq}).
The results for each diagram are summarized in the Table~\ref{tab:Tab3}. 
The appearance of the trigonometric cosine functions in some of the terms is due to the fact that the corresponding loop diagrams contain  
unequal numbers of advanced/retarded propagators. 
Each diagram in the respective class is then complex, but the sum of all complex conjugate diagrams is real and proportional to the cosine.
Even numbers in the argument of the cosine are equal to the difference between the number of propagators in each side. 
Because our series contain only cosine functions they might seem different in comparison to Refs.~[\onlinecite{Hikami1984a,Hikami1984b,Chakravarty1986,Hikami1993}].
There is no discrepancy though, since these authors use extensively the addition formulae of trigonometric functions, cf. especially Ref.~[\onlinecite{Hikami1984b}].

Diagrams in every order organize in distinct subgroups by the topology of the diagrams. The sum of all terms in a subgroup is a real function. 
We summarize the results for every subgroup in expansion of the second moment of the position operator $\sum_r r^2_\mu P^{}_{r,0}$ in Table~\ref{tab:Tab3}
and for the corresponding normalization $\sum_r P^{}_{r,0}$ in Table~\ref{tab:Tab4}. 
Each subgroup provides in the $n$th order of $\sum_r r^2_\mu P^{}_{r,0}$ for a contribution 
\begin{equation}
\label{eq:CFsm}
{\cal R}^{(n)}_i = \frac{1}{4\pi}\frac{1}{E^2_g} \sum_s (2X^{}_s)^{2n+2} \cos\left[({\cal N}^+_i-{\cal N}^-_i)\phi^{}_s\right]{\cal C}^{(n)}_{2,i},
\end{equation}
where $ X^2_s = E^2_g [\eta^2_s(E) + \rho^2_s(E)]$, $\phi^{}_s = {\rm arctan}\left[\frac{\rho^{}_s}{\eta^{}_s}\right]$, 
and ${\cal N}^\pm_i$ is the number of advanced (+) and retarder ($-$) propagators in the propagator, obeying ${\cal N}^+_i+{\cal N}^-_i = 2n$, 
and $i$ is the index of the subgroup as it appears in Tables~\ref{tab:Tab3} and Table~\ref{tab:Tab4}. 
The corresponding expression for the normalization $\sum_r P^{}_{r,0}$ reads
\begin{equation}
\label{eq:CFn}
{\cal Z}^{(n)}_i = \frac{k^2}{4\pi}\frac{1}{E^2_g} \sum_s (2X^{}_s)^{2n+2} \cos\left[({\cal N}^+_i-{\cal N}^-_i)\phi^{}_s\right]{\cal C}^{(n)}_{0,i}.
\end{equation}
The factors ${\cal N}^+_i-{\cal N}^-_i$ can be read off directly from the Fig.~\ref{fig:Fig6} giving 
${\cal N}^+_i-{\cal N}^-_i = 0$ for $i=1,2,3,4,6,14,15,16,17$, ${\cal N}^+_i-{\cal N}^-_i = 2$ for $i=5,10,11,12,13$, and finally ${\cal N}^+_i-{\cal N}^-_i = 4$ for $i=7,8,9$.

The combinatorial factors ${\cal C}^{(n)}_{2,i}$ for each subgroup represent rational numbers which appear as the result of integration in the position space 
(all variables rescaled in units of $k$)
\begin{equation}
{\cal C}^{(n)}_{2,i} = \int \frac{d^2r}{2\pi} ~ r^2e^{-r^2}
\underbrace{
\int\frac{d^2x}{\pi}\cdots \int \frac{d^2y}{\pi}\cdots ,
}_{\rm n \,\, integrals}
\end{equation}
where the additional factor $1/2$ in the very first integral is due to the angular average $\langle r^2_\mu\rangle^{}_\varphi = 1/2\langle r^2\rangle^{}_\varphi$. 
These integral chains are unique for each diagram.
Correspondingly, the combinatorial factors ${\cal C}^{(n)}_{0,i}$ for the normalization are defined as 
\begin{equation}
{\cal C}^{(n)}_{0,i} = \int \frac{d^2r}{\pi} ~ e^{-r^2}
\underbrace{
\int\frac{d^2x}{\pi}\cdots \int \frac{d^2y}{\pi}\cdots .
}_{\rm n \,\, integrals}
\end{equation}
There is a simple relation between  ${\cal C}^{(n)}_{2,i}$ and ${\cal C}^{(n)}_{0,i}$
\begin{equation}
{\cal C}^{(n)}_{0,i} = 2\chi^{}_i {\cal C}^{(n)}_{2,i},
\end{equation}
where $\chi^{}_i$ are taken from the Table~\ref{tab:Tab1}.
The combinatorial factors ${\cal C}^{(n)}_{2,i}$ and ${\cal C}^{(n)}_{0,i}$ for every term are summarized in Tables~\ref{tab:Tab3} and Table~\ref{tab:Tab4} respectively.

\begin{table}
\begin{tabular}{ccccccccccc}
No. &  Diagram  &  Amplitude & \hspace{2mm} &  No.  & {\rm Diagram } &  Amplitude & \hspace{2mm} & No. & {\rm Diagram} & Amplitude   \\
\\
1
&
\includegraphics[height=3mm]{D0.eps} 
&  
$\frac{1}{2}$
&
&
6
& 
\includegraphics[height=3mm]{D3t1.eps} 
&
$\frac{2}{9}$
&
&
12
&
\includegraphics[height=3mm]{D3t7.eps}
&
$\frac{3}{8}$
\\
\\
2
&
\includegraphics[height=3mm]{D1.eps}
& 
$1$
&
&
7
&
\includegraphics[height=3mm]{D3t2.eps} 
&
$\frac{4}{9}$
&
&
13
&
\includegraphics[height=7mm]{D3t8.eps} 
&
$\frac{4}{3}$
\\
\\
3
&
\includegraphics[height=3mm]{D2t1.eps} 
&
$\frac{3}{2}$
&
&
8
&
\includegraphics[height=3mm]{D3t3.eps}
&
$\frac{3}{8}$
&
&
14
&
\includegraphics[height=3mm]{D3t9.eps} 
&
$2$
\\
\\
4
&
\includegraphics[height=3mm]{D2t2.eps} 
&
$\frac{1}{2}$
&
&
9
&
\includegraphics[height=7mm]{D3t4.eps}
&
$\frac{10}{9}$
&
&
15
&
\includegraphics[height=3mm]{D3t10.eps} 
&
$\frac{2}{3}$
\\
\\
5
&
\includegraphics[height=3mm]{D2t3.eps} 
&
$\frac{3}{4}$
&
&
10
&
\includegraphics[height=3mm]{D3t5.eps}
&
$\frac{5}{9}$
&
&
16
&
\includegraphics[height=3mm]{D3t11.eps}
&
$\frac{3}{2}$
\\
\\
&
&
&
&
11
&
\includegraphics[height=7mm]{D3t6.eps}
&
$\frac{5}{2}$
&
&
17
&
\includegraphics[height=3mm]{D3t12.eps}
&
$\frac{1}{4}$
\end{tabular}
\caption{Amplitudes ${\cal C}^{(n)}_{2,i}$ of every term of $\sum_r r^2_\mu P^{}_{r,0}$ up to the order $g^3$, $i$ is the number of the diagram. 
The net contribution from each diagram is acquired by inserting the elements of the Table into Eq.~(\ref{eq:CFsm}).
In every order, the ladder diagrams weight substantially heavier as all other non-oscillating terms (no. 4,6,15,16,17) combined.}
\label{tab:Tab3} 
\end{table}

\begin{table}
\begin{tabular}{ccccccccccc}
No. &  Diagram  &  Amplitude & \hspace{2mm} &  No.  & {\rm Diagram } &  Amplitude & \hspace{2mm} & No. & {\rm Diagram} & Amplitude   \\
\\
1
&
\includegraphics[height=3mm]{D0.eps} 
&  
$1$
&
&
6
& 
\includegraphics[height=3mm]{D3t1.eps} 
&
$\frac{1}{3}$
&
&
12
&
\includegraphics[height=3mm]{D3t7.eps}
&
$\frac{1}{2}$
\\
\\
2
&
\includegraphics[height=3mm]{D1.eps}
& 
$1$
&
&
7
&
\includegraphics[height=3mm]{D3t2.eps} 
&
$\frac{2}{3}$
&
&
13
&
\includegraphics[height=7mm]{D3t8.eps} 
&
$\frac{4}{3}$
\\
\\
3
&
\includegraphics[height=3mm]{D2t1.eps} 
&
$1$
&
&
8
&
\includegraphics[height=3mm]{D3t3.eps}
&
$\frac{1}{2}$
&
&
14
&
\includegraphics[height=3mm]{D3t9.eps} 
&
$1$
\\
\\
4
&
\includegraphics[height=3mm]{D2t2.eps} 
&
$\frac{1}{2}$
&
&
9
&
\includegraphics[height=7mm]{D3t4.eps}
&
$\frac{4}{3}$
&
&
15
&
\includegraphics[height=3mm]{D3t10.eps} 
&
$\frac{2}{3}$
\\
\\
5
&
\includegraphics[height=3mm]{D2t3.eps} 
&
1
&
&
10
&
\includegraphics[height=3mm]{D3t5.eps}
&
$\frac{2}{3}$
&
&
16
&
\includegraphics[height=3mm]{D3t11.eps}
&
$1$
\\
\\
&
&
&
&
11
&
\includegraphics[height=7mm]{D3t6.eps}
&
$2$
&
&
17
&
\includegraphics[height=3mm]{D3t12.eps}
&
$\frac{1}{4}$
\end{tabular}
\caption{Amplitudes ${\cal C}^{(n)}_{0,i}$ of every term of $\sum_r P^{}_{r,0}$  up to the order $g^3$, $i$ is the number of the diagram. 
The net contribution for the respective expansion order from each diagrams is acquired by inserting the elements of the Table into Eq.~(\ref{eq:CFn}).}
\label{tab:Tab4} 
\end{table}

The inspection of the Gaussian exponents in Table~\ref{tab:Tab1} suggests that the dominant contribution on large scales in every order of expansion arises from the 
so-called ladder diagrams, i.e. diagrams no. 1,2,3, 14 etc. According to Table~\ref{tab:Tab3} they provide more weight 
in every order of expansion of $\sum_r r^2_\mu P^{}_{r0}$ than all other non-oscillating diagrams (without cosine) combined. 
In particular the so-called fan diagrams, e.g. no. 4, 17, etc. contribute the least in every order of expansion. 
This is the ultimate reason for neglecting the contributions from the latter. Four lowest order ladder diagrams are 
\begin{eqnarray}
\includegraphics[height=3mm]{D0.eps} = \frac{1}{4E^2_g}\left(\frac{k^2}{\pi}\right)^2\sum_{s=\pm}(2X^{}_s)^2\exp\left[-k^2r^2\right],  & & 
\includegraphics[height=3mm]{D1.eps} = \frac{1}{4E^2_g}\left(\frac{k^2}{\pi}\right)^2\sum_{s=\pm}~ \frac{(2X^{}_s)^4}{2}\exp\left[-\frac{k^2r^2}{2}\right],\\
\includegraphics[height=3mm]{D2t1.eps} = \frac{1}{4E^2_g}\left(\frac{k^2}{\pi}\right)^2\sum_{s=\pm}~ \frac{(2X^{}_s)^6}{3}\exp\left[-\frac{k^2r^2}{3}\right], && 
\includegraphics[height=3mm]{D3t9.eps} = \frac{1}{4E^2_g}\left(\frac{k^2}{\pi}\right)^2\sum_{s=\pm}~ \frac{(2X^{}_s)^8}{4}\exp\left[-\frac{k^2r^2}{4}\right],
\end{eqnarray}
which suggests the following approximate expression for the two-particles propagator:
\begin{equation}
P^{\rm lad}_{r0}(E) \approx  \frac{1}{4E^2_g}\left(\frac{k^2}{\pi}\right)^2\sum_{s=\pm} ~ \sum^\infty_{n=0}\frac{(2X^{}_s)^{2n+2}}{n+1}\exp\left[-\frac{k^2r^2}{n+1}\right].
\end{equation}
The dominant oscillating correction comprises the menora diagrams no. 5, 7 etc., Table~\ref{tab:Tab1}. 
The contribution from the diagrams 5 to the two-particles propagator reads 
\begin{equation}
\includegraphics[height=3mm]{D2t3.eps}  = \frac{1}{4E^2_g} \left(\frac{k^2}{\pi}\right)^2\sum_{s=\pm} ~\frac{2}{3} \exp\left[-\frac{2}{3}k^2r^2\right]\left(2X^{}_s\right)^6\cos2\phi^{}_s,
\end{equation}
from the diagrams 7
\begin{equation}
\includegraphics[height=3mm]{D3t2.eps}  = \frac{1}{4E^2_g} \left(\frac{k^2}{\pi}\right)^2\sum_{s=\pm} ~\frac{2}{4} \exp\left[-\frac{3}{4}k^2r^2\right]\left(2X^{}_s\right)^8\cos4\phi^{}_s,
\end{equation}
and so on. 
The full series accounting for the contribution from the menora channel then becomes
\begin{equation}
\label{eq:MenCl}
P^{\rm men}_{r0}(E) = \frac{2}{4E^2_g} \left(\frac{k^2}{\pi}\right)^2\sum_{s=\pm}\sum^{\infty}_{n=2}\frac{(2X^{}_s)^{2n+2}}{n+1}\cos[(2n-2)\phi^{}_s]\exp\left[-\frac{n}{n+1}k^2r^2\right].
\end{equation}
Both ladder and menora channels combined represent a reasonable exact large scale asymptotics for the dressed two-particles propagator.

\section{Details of the derivation of Eq.~(\ref{eq:DiffEq})}
\label{app:ScalEq}

We begin with Eq.~(\ref{eq:InMSqD2}) and use the fluctuations $\langle\delta f^{(i)}\rangle$ as the generators of the time evolution of 
the mean squared displacement at small positive/negative times $\Delta t$:
\begin{equation}
\label{eq:InMSqD3}
k^2\langle r^2_{\mu}(\pm \Delta t)\rangle \approx 
\frac{\displaystyle \langle f^{(2)}\rangle + \sum^{}_{s=\pm}\int^\infty_{-\infty} d\nu^{}_s\left[e^{\pm i\nu^{}_sE^{}_g\Delta t}-1\right] \frac{N}{2}(2X^{}_s)^{2N}}
{\displaystyle \langle f^{(0)}\rangle + \sum^{}_{s=\pm}\int^\infty_{-\infty} d\nu^{}_s\left[e^{\pm i\nu^{}_sE^{}_g\Delta t}-1\right](2X^{}_s)^{2N}},
\end{equation}
where we carry out the Fourier transform in the energy units specific for the time representation of the Wegner's propagator Eq.~(\ref{eq:WGFT}).
Obviously, the case $\Delta t=0$ restores Eq.~(\ref{eq:InMSqD3}). For small $(e^{\pm i\nu^{}_sE^{}_g\Delta t}-1)$ we have 
\begin{eqnarray}
k^2[\langle r^2_{\mu}(\pm \Delta t)\rangle - \langle r^2_{\mu}(0)\rangle] \approx \frac{1}{\langle f^{(0)}\rangle}
\sum^{}_{s=\pm}\int^\infty_{-\infty} d\nu^{}_s\left[e^{\pm i\nu^{}_sE^{}_g\Delta t}-1\right] (2X^{}_s)^{2N} \left[\frac{N}{2} - k^2 \langle r^2_{\mu}(0)\rangle \right].
\end{eqnarray}
Summing both sides yields 
\begin{eqnarray}
k^2[\langle r^2_{\mu}(\Delta t)\rangle - 2\langle r^2_{\mu}(0)\rangle + \langle r^2_{\mu}(-\Delta t)\rangle] &\approx&
\frac{2}{\langle f^{(0)}\rangle} \sum^{}_{s=\pm}\int^\infty_{-\infty} d\nu^{}_s\left[\cos(\nu^{}_sE^{}_g\Delta t)-1\right] \\
&\times&(2X^{}_s)^{2N}\left[\frac{N}{2} - k^2 \langle r^2_{\mu}(0)\rangle \right].
\end{eqnarray}
Taking the limit $\Delta t\to 0$ and using the self-consistent approximation yields Eq.~(\ref{eq:DiffEq}).

\end{document}